\documentclass[fleqn,usenatbib]{mnras}
\DeclareUnicodeCharacter{2212}{-}

\usepackage{newtxtext,newtxmath}
\usepackage{soul}

\usepackage[T1]{fontenc}

\DeclareRobustCommand{\VAN}[3]{#2}
\let\VANthebibliography\thebibliography
\def\thebibliography{\DeclareRobustCommand{\VAN}[3]{##3}\VANthebibliography}

\usepackage{graphicx}	
\usepackage{amsmath}	
\usepackage{xcolor}

\newcommand{\cielo}{{\sc cielo}}

\title[Evolution of DM halo shape]{Redshift evolution of the dark matter haloes shapes}

\author[P. Cataldi et al.]{
P. Cataldi$^{1}$\thanks{Contact e-mail: pcataldi@iafe.uba.ar},
S. E. Pedrosa$^{1}$,
P. B. Tissera$^{2,3}$,
M. C. Artale$^{4,5,6}$,
N. D. Padilla$^{7}$, 
\newauthor
R. Dominguez-Tenreiro$^{8,9}$,
L. Bignone$^{1}$,
R. Gonzalez$^{3}$
and
L.J. Pellizza $^{1}$
\\
$^{1}$Instituto de Astronom\'{\i}a y F\'{\i}sica del Espacio, CONICET-UBA, Casilla de Correos 67, Suc. 28, 1428, Buenos Aires, Argentina\\
$^{2}$ Instituto de Astrof\'{i}sica, Pontificia Universidad Cat\'olica de Chile, Av. Vicuña Mackenna 4860, Santiago, Chile.\\
$^{3}$Centro de Astro-Ingenier\'ia, Pontificia Universidad Cat\'olica de Chile, Av. Vicu\~na Mackenna 4860, Santiago, Chile.\\
$^{4}$ Physics and Astronomy Department Galileo Galilei, University of Padova, Vicolo dell’Osservatorio 3, I-35122, Padova, Italy\\
$^{5}$ INFN - Padova, Via Marzolo 8, I-35131 Padova, Italy\\
$^{6}$ Department of Physics and Astronomy, Purdue University, 525 Northwestern Avenue, West Lafayette, IN 47907, USA \\
$^{7}$ Instituto de Astronom\'{\i}a Te\'orica y Experimental, UNC-CONICET, Laprida 854, X5000BGR C\'ordoba, Argentina \\
$^{8}$ Departamento de F\'{\i}sica Te\'orica, Universidad Aut\'onoma de Madrid, E-28049 Cantoblanco, Madrid, Spain \\
$^{9}$ Centro de Investigaci\'on Avanzada en F\'{\i}sica Fundamental, Universidad Aut\'onoma de Madrid, E-28049 Cantoblanco, Madrid, Spain \\
}

\date{Accepted XXX. Received YYY; in original form ZZZ}

\pubyear{2022}

\begin{document}
\label{firstpage}
\pagerange{\pageref{firstpage}--\pageref{lastpage}}
\maketitle

\begin{abstract}

In this work, we aim at investigating the morphology evolution of Milky Way mass-like dark matter haloes  selected from the  \cielo{}  and {\sc IllustrisTNG} Projects. The connection between halo shapes and their environment has been studied in previous works at z=0 but their connection remains yet to be fully understood. We focus on the evolution across cosmic time of the halo shapes and the relation with the infalling material, using hydrodynamical simulations. Our findings show that haloes tend to be more triaxial at earlier times as a consequence of stronger accretion in the direction of the filaments. As the haloes evolve towards a dominant isotropic accretion mode and relaxation, their shape at 20 percent of the virial mass becomes more spherical. In agreement with previous results, baryons have an important effect within the inner regions of the haloes, driving them from triaxial to rounder shapes. We also find a correlation between the strength of the quadrupole infalling mode and the degree of ellipticity of the haloes: as the filament strength decreases steadily with redshift, the haloes became more spherical and less elliptical.    
\end{abstract}




\section{Introduction}

In the current cosmological paradigm $\Lambda$-CDM model, dark matter (DM) and dark energy are the main ingredients that drive the formation and evolution of cosmic structures. In particular, DM haloes grow hierarchically and continuously by successive mergers and accretion \citep{Zeldovich1970}, embedded within the filamentary structures of the cosmic web  \citep[e.g.][]{White1978,Peebles1980,Ghigna1998,Springel2008}.

The characteristic structures of DM haloes have been studied extensively using cosmological simulations. Early N-body simulations show that DM haloes can be described by a universal radial density profile (NFW, \citealt{Navarro1997}), while their shape tends to be triaxial and prolate in the inner regions \citep[e.g.][]{Frenk1988,Jing2002,Allgood2005,Stadel2009}. Cosmological hydrodynamical simulations, which account for the DM assembly together with galaxy formation have proved to be of great help to investigate the impact of baryons on different properties of the haloes \citep{Tissera2010}. In fact, baryons and the physics processes associated to their evolution can modify the shape of DM haloes sphericalizing them \citep[][]{Tissera1998, Kazantzidis2004,Zemp2012,Zhu2017,Chua2019,Cataldi2020}, change the distribution of the DM in the inner regions \citep[][]{Governato2012,DiCintio2014,Artale2018} and change their specific angular momentum \citep[][]{Pedrosa2010,Zavala2016,Lagos2017}.

Regarding the mass density profile, \citet{Tollet2016} investigated the properties of haloes with virial masses in the range of $\mathrm{10^{10}-10^{12} M_{\odot}}$, finding that in the central regions of the halo, the inner DM density slope depends on the stellar-to-halo mass ratio at all analysed redshifts, in agreement with \citet{DiCintio2014} which investigate this dependence at $z=0$. 

More recently, \citet{Artale2018}, inspecting the mass accretion history (MAH) of the DM haloes selected from a cosmological hydrodynamical simulation, showed that they assemble earlier that their dark matter only (DMo) counterparts. This change in formation history was explained since baryons make haloes more concentrated and in turn more massive than their DMo counterparts, reporting a close connection between the MAH, the amount of baryons, and the evolution of the DM density profiles.

The spherical collapse and gaussian random fields model of halo formation \citep{Press1974} needs a extending as the mass accretion on the haloes falls along a preferential direction  \citep[mostly along the filaments][]{Zeldovich1970} and tends to be clumpy. Given this preferential direction in the accretion along filaments, DM haloes show mostly non-spherical symmetry, especially if their relaxation times are no larger than the time between mergers or accretion events. In contrast, early-formed objects were weakly connected to their environment and were highly relaxed \citep{Gouin2021}. 

\citet{Ludlow2014} and \citet{Bonamigo2015} studying DMo simulations, used the dimensionless parameter peak height, $\nu (M,z)$, to characterise the shape of haloes within a wide range of masses and redshifts. They found that DM haloes are triaxial with a tendency to be prolate. In particular, more massive objects are less spherical than low mass haloes, essentially because high mass haloes formed later on
\citep{Despali2017}. This increase in triaxiality, correlates both with mass and with redshift because haloes seem to be affected by the direction of the last major merger accreted along the filaments around them \citep{Jing2002,Allgood2005,Vega-Ferrero2017}. Following the same approach as \citet{Bonamigo2015}, \citet{Vega-Ferrero2017} found that minor-to-major axis ratio can be expressed by a universal function in terms of $\nu (M,z)$.

It has been reported \citep[][]{Oguri2005,Giocoli2012a, Giocoli2012b,Wojtak2013,Lau2021} a close connection between the triaxiality of DM haloes and the cluster mass and concentration and the inner slope of the DM density profile, with consequences on the strong lensing cross-sections. Satellite galaxies are preferentially accreted along filaments \citep{Libeskind2014,Tempel2015}. Because infall is driven by the surrounding large-scale structure, we expect a significant correlation between the halo shapes and their environment \citep[][reference within]{Vera-Ciro2012}. \citet{Governato2012} found that haloes tend to point their minor axes perpendicular to the infall (filament) direction. Additionally, subhalos are predominantly accreted along the major axis of the host halo, and the alignment increases with the host halo mass \citep{Kang2015}.

At later times the cross-section of the filaments becomes larger than the typical size of the MW-mass haloes and, as a result, accretion turns more isotropic and the objects evolve into a more oblate configuration. Interestingly, haloes retain memory of their structure at earlier times \citep{Vera-Ciro2012}. This is imprinted in their present-day shape dependency with  radius, which changes from typically prolate in the inner (earlier collapsed) regions to a triaxial in the outskirts (corresponding to the shells that have last collapsed  and are now at about the virial radius).

In the case of MW-like galaxies, \citet{Shao2021} investigated how the disc of satellite galaxies can be used to infer the orientation and some aspects of the formation history of the Galactic DM halo, using the \textsc{EAGLE} simulation. These authors found that the normal to the common orbital plane of satellites, as well as the central stellar disc, is well aligned with the minor axis of the DM host halo. Also, \citet{Shao2021} found that the DM halo of each of their MW-analogue is "twisted" such that the orientation of the outer halo is perpendicular to that of the inner halo. This occurs because the inner halo is aligned with the central disc, whereas the outer halo is nearly perpendicular to the stellar disc, with a tight alignment towards the filamentary network along which mass is accreted . 

Observational studies based on using X-ray data \citep{Fabricant1984, Buote1996,Kawahara2010,Lau2013}, Sunyaev Zel'dovich \citep{Sayers2011} and strong and weak gravitational lensing methods \citep{Soucail1987, Evans2009,Oguri2010,Oguri2012} indicate that cluster-size DM haloes are often not spherical. However, the observational determination of the shapes of DM haloes is quite challenging. Few studies attempted to infer the shape and orientation of the galactic DM halo. Preferably dynamical tracers at large radii are to be used, in which many cases, by definition are rare \citep{Vera-Ciro2012}. Dynamical tracers such as kinematics and morphology of the HI layer have been used to impose constraints on the halo morphology \citep{Becquaert1997,Swaters1997}, also the temperature profile of X-ray isophotes \citep{Buote1998,Buote2002}, gravitational lensing \citep{Hoekstra2004} and the spatial distribution of galaxies within groups \citep{Paz2006,Robotham2008}. The general trend of all these studies is that haloes tend to be roughly oblate, with the smallest axis pointing perpendicular to the symmetry plane defined by the stellar component. 

In the case of the MW, the shape constraints often rely on the kinematics of stars, which include the proper motions of hypervelocity stars \citep{Gnedin2005} or the dynamics of stellar streams \citep{Koposov2010}. These studies show a nearly spherical Galactic halo \citep[][]{Ibata2001,Law2005,Law2009,Law2010, Bovy2016, Malhan2019}, in agreement with numerical studies \citep[e.g.][]{Chua2019,Cataldi2020}.

Studying the shape, the MAH and the concentration of haloes provides an opportunity to learn about individual growth histories and the connection between them and the properties of their host galaxy. \citep{Drakos2019}. Measurements of structural properties for large, well-defined samples of haloes may also provide new cosmological tests \citep[see e.g.][for discussion]{Taylor2011}.

In this paper, we use  a set of haloes identified from the \cielo~ and {\sc  IllustrisTNG} projects to deepen the impact of different sub-grid models and different cosmic environments might have on the DM halo morphologies. These results extend and strengthen previous studies on halo shapes \citep{Cataldi2020, Cataldi2022}, by including now the temporal evolution of the analysed properties.

The paper is structured as follows. Section \ref{sec:sim} reviews the simulations setup and sample selection. Section \ref{sec:res} presents the results on the evolution of DM structure divided into: subsection \ref{sec:MAH} focuses on analyzing the mass accretion history and halo size evolution, subsection \ref{sec:Morph} analyze the shape profiles evolution and their dependence with merger events, and  subsection \ref{sec:Infall}  studies the impact of the infall matter configurations on the shape analysis. Finally, conclusions  are summarized in Section \ref{sec:Conc}.

\section{Simulations}
\label{sec:sim}

Here we use two simulation suites, namely the \cielo{} and {\sc  IllustrisTNG} \citep[][]{Nelson2019} simulations which were run with different prescriptions for the baryonic processes involved in galaxy formation. We summarize the main features of each case.

\subsubsection{\cielo~ simulations}

The Chemo-dynamIcal propertiEs of gaLaxies and the cOsmic web, \cielo{},  is a project aimed to study the formation of galaxies in different environments, with virial mass haloes within the range $\mathrm{M_{200} = 10 ^{10}− 10^{12} M_{\odot}}$ \citep{Rodriguez2022}. It also includes two Local Group (LG) analogues. The \cielo~ simulations assume a $\Lambda$-CDM universe model with a cosmology consistent with \citet{Planck2013}, given by $\Omega _{0} = 0.317$, $\Omega_ {\Lambda} = 0.6825$, $\Omega _{B} = 0.049$ and $h = 0.6711$. 

\cielo~ was run with a version of \textsc{GADGET-3} based on \textsc{GADGET-2} \citep{springel2003,springel2005}.  It includes a multi-phase model for the gas component metal-dependant cooling, star formation and  energy feedback Type II and Type Ia Supernovae (SNeII and SNeIa, respectively), as described by \citet{scan2005} and \citet{scan2006}. The \cielo~  simulations assume an Initial Mass Function of \citet{Chabrier2003}. The chemical evolution model included  follows the enrichment  by SNII and SNIa types, keeping track of 12 different chemical elements \citep[][]{mosco2001}. This version of \textsc{GADGET-3} has been previously used by \citet{Pedrosa2015} to study the mass-size relation and specific angular momentum content of galaxies, and by \citet{Tissera2016a,Tissera2016b} to investigate the origin of the metallicity gradients of the gas-phase components and stellar population of galaxies in the {Fenix} simulation. 

The initial conditions of the \cielo~ simulations were taken from a DMo run of a cosmological periodic cubic box of side length $\mathrm{L = 100 Mpc \mathit{h^{ −1}}}$. The MUSIC code \citep{Hahn2011}, which computes multi-scale cosmological initial conditions under different approximations and transfer functions, was applied to extract the object and increase the numerical resolution. A first set of 20 LG analogues were initially selected and two pairs of them (LG1 and LG2) were chosen by imposing constraints on the relative velocity, separation and the mass of the DM haloes \citep[see,][]{Rodriguez2022}.The two LG selected were re-run with a DM particle resolution of $\mathrm{m_{dm} = 1.2 \times 10^{6} M_{\odot} \, \mathit{h^{ −1}}}$. Baryons were added with an initial gas mass of $\mathrm{m_{baryon}=2.0 \times 10^5  M_{\odot} \, \mathit{h^{ −1}}}$.
\subsubsection{{\sc IllustrisTNG} simulations}

\textit{The Next Generation Illustris Simulation } ({\sc IllustrisTNG}) suite is a set of cosmological simulations that were run with the \textsc{AREPO} code \citep[][]{Weinberger2020}. The {\sc  IllustrisTNG} simulation follows an updated model of galaxy formation based on the results from the original Illustris simulation, which includes subgrid models to account for different baryonic processes such as star formation, stellar feedback, gas cooling and AGN feedback \citep[see][]{Weinberger2017,Pillepich2018, Nelson2019}. The updated model of {\sc  IllustrisTNG} includes cosmic magnetic fields and adopts a cosmology consistent with  \citet{Planck2016}, given by $\Omega _{m}=0.3809$, $\Omega _{\Lambda}=0.6911$, $\Omega _{b}=0.0486$, $\sigma_{8}=0.8159$, $n_{s}=0.9667$, $h=0.6774$.
 
The initial conditions of {\sc IllustrisTNG} were generated with the Zel'dovich approximation \citep{Zeldovich1970}. Here we use the highest resolution available for {\sc  IllustrisTNG 50} (hereafter, TNG50). The TNG50 consists of a periodic box size of $\mathrm{35 Mpc\,\mathit{h^{ −1}}}$. TNG50 contains $2160^3$ DM particles and the same initial number of gas cells. The mass of the DM particles is uniform, $\mathrm{m_{dm}=3.0 \times 10^5  M_{\odot} \, \mathit{h^{ −1}}}$, and the average mass of the gas cells (and stellar particles) is $\mathrm{m_{baryon}=5.8 \times 10^4  M_{\odot} \, \mathit{h^{ −1}}}$. 

\subsection{Haloes selection}

In \cielo~simulations, galaxies are identified by using a Friends-of-Friends and SUBFIND algorithms. Among them, we select for this study the most massive central galaxies from simulated LG1 and LG2. In particular, LG1 was previously analysed by \citet{Rodriguez2022} to study the evolution of infalling disc satellites and by \citet{Tapia2022} to analyse the metallicity gradients of the central galaxies. 

In the case of TNG50, we select haloes from the simulated box, restricting the stellar masses of the host galaxy of the haloes to the range of $4 \times 10^{9} <\mathrm{M_{star}}/{\rm M_{\odot}}< 6 \times 10^{10}$, with a star formation rate and metallicity above zero. In this work, we use the virial radius, $\mathrm{r_{200}}$, and the virial mass, $\rm M_{200}$, as the radius and mass within a sphere containing $\sim 200$ times the cosmic critical matter density at the corresponding redshift.

From the 517 haloes fulfilling the aforementioned conditions in TNG50, we remove those with a recent major merger. For this purpose, we select galaxies with minor mergers since $z=2$, with the stellar mass ratio $\mu<1/4$. This constraint follows recent studies of \citet{Helmi2018} and \citet{Belokurov2018} who inferred  that the Galaxy has undergone the last major merger event around at $z=1-2$ by using data from \textit{Gaia} mission \citep[][]{Gaia2018}. From our main sample, we found 15 MW-like haloes that fulfill all selection criteria adopted. For the TNG50 selection, we do not impose environment constraints such as physical separation or relative velocity on other nearby simulated galaxies.

Since the main purpose of this work is to follow in time the evolution of halo morphologies, we made use of the subhalo merger tree catalogue available in each simulation. The \cielo~ simulations have a merger tree computed using the \textsc{MergerTree} routine of the \textsc{Amiga Halo Finder} \citep{Knollmann2009} and in the case of TNG50, the {\sc  IllustrisTNG} simulation data base provides a merger trees computed with SubLink algorithm \citep{Rodriguez-Gomez2015}, which we use in order to select and follow back the main branch of subhaloes chosen at $z=0$.

\begin{table}
	\centering
	\caption{An overview of the main characteristics of the selected \cielo{} and TNG50 haloes at $z=0$. From left to right, we show the halo IDs, the total virial mass ($\mathrm{M_{200}^{tot}}$), the total stellar and DM virial mass, ($\mathrm{M_{200}^{star}}$, $\mathrm{M_{200}^{DM}}$) and the virial radius ($\mathrm{r_{200}}$). In bold, the haloes with a recent major mass accretion.}
	\label{Tab:simu}
	\begin{tabular}{lcccr} 
		\hline
		\cielo{} & $\mathrm{M_{200}^{tot}[M_{\odot}/\mathit{h}]}$ & $\mathrm{M_{200}^{star}[M_{\odot}/\mathit{h}]}$  & $\mathrm{M_{200}^{DM}[M_{\odot}/\mathit{h}]}$& $\mathrm{r_{200}[kpc/\mathit{h}]}$ \\
		\hline
		h4337 & $\mathrm{9.4 \times 10^{11}}$ & $\mathrm{4.0 \times 10^{10}}$ & $\mathrm{8.7 \times 10^{11}}$ & $\mathrm{159.9}$\\
		\textbf{h4469} & $\mathrm{3.6 \times 10^{11}}$ & $\mathrm{6.1 \times 10^{9} }$  & $\mathrm{3.4 \times 10 ^{11}}$ & $\mathrm{137.0}$\\
		h87 & $\mathrm{3.6 \times 10^{11}}$ & $\mathrm{2.8 \times 10^{9}}$  & $\mathrm{3.4 \times 10^{11}}$   & $\mathrm{118.8}$ \\
		h115  & $\mathrm{2.0 \times 10^{11}}$ & $\mathrm{3.2 \times 10^{9}}$   & $\mathrm{1.9 \times 10^{11}}$  & $\mathrm{97.0}$\\
		\hline
    	TNG50 & $\mathrm{M_{200}^{tot}[M_{\odot}/\mathit{h}]}$ & $\mathrm{M_{200}^{star}[M_{\odot}/\mathit{h}]}$  & $\mathrm{M_{200}^{DM}[M_{\odot}/\mathit{h}]}$& $\mathrm{r_{200}[kpc/\mathit{h}]}$\\
        \hline
		\textbf{h476266} & $\mathrm{1.1 \times 10^{12}}$ & $\mathrm{5.0 \times 10^{10}}$  & $\mathrm{9.4 \times 10^{11}}$   & $\mathrm{161.3}$ \\
		h533590  & $\mathrm{5.6 \times 10^{11}}$ & $\mathrm{2.3 \times 10^{10}}$   & $\mathrm{4.9 \times 10^{11}}$ & $\mathrm{125.0}$\\
		h593480  & $\mathrm{4.7 \times 10^{11}}$ & $\mathrm{2.1 \times 10^{10}}$   & $\mathrm{4.1 \times 10^{11}}$ & $\mathrm{112.6}$\\
		h631558  & $\mathrm{2.9 \times 10^{11}}$ & $\mathrm{1.2 \times 10^{10}}$   & $\mathrm{2.5 \times 10^{11}}$ & $\mathrm{95.6}$\\
		h649627  & $\mathrm{2.3 \times 10^{11}}$ & $\mathrm{8.0 \times 10^{9}}$   & $\mathrm{2.0 \times 10^{11}}$ & $\mathrm{88.7}$\\
		h656142  & $\mathrm{2.4 \times 10^{11}}$ & $\mathrm{6.8 \times 10^{9}}$   & $\mathrm{2.2 \times 10^{11}}$ & $\mathrm{89.1}$\\  
        \hline
	\end{tabular}
\end{table}

Therefore the haloes from \cielo{} simulations analysed were h4337 (LG1), h87 (LG2) and h115 (LG2). In the case of TNG50, we opt to analyse one-third of the selection, 5 of them with halo IDs h533690, h593480, h631558, h649627 and h656142, as we want to focus on the detailed individual changes of their structure evolution. Table \ref{Tab:simu} summarises the main properties of the MW-like analogues. We kept for comparison analysis, two haloes without constraints in merger activity. From \cielo{}, LG1-h4469, which had a major mass accretion due to a closed interaction at $z=0.38$ and from TNG50, h476266, which presents intermediate merger events since $z=2$. Both haloes are indicated in Table \ref{Tab:simu} with boldface. 

Although the nearly identical cosmologies between TNG50 and \cielo, we chose to show the quantities in terms of $h$ to compare results between simulations.

\section{Results}
\label{sec:res}

\subsection{The mass density profile evolution}
\label{sec:MAH}

\begin{figure}
\centering
\includegraphics[width=0.512\columnwidth]{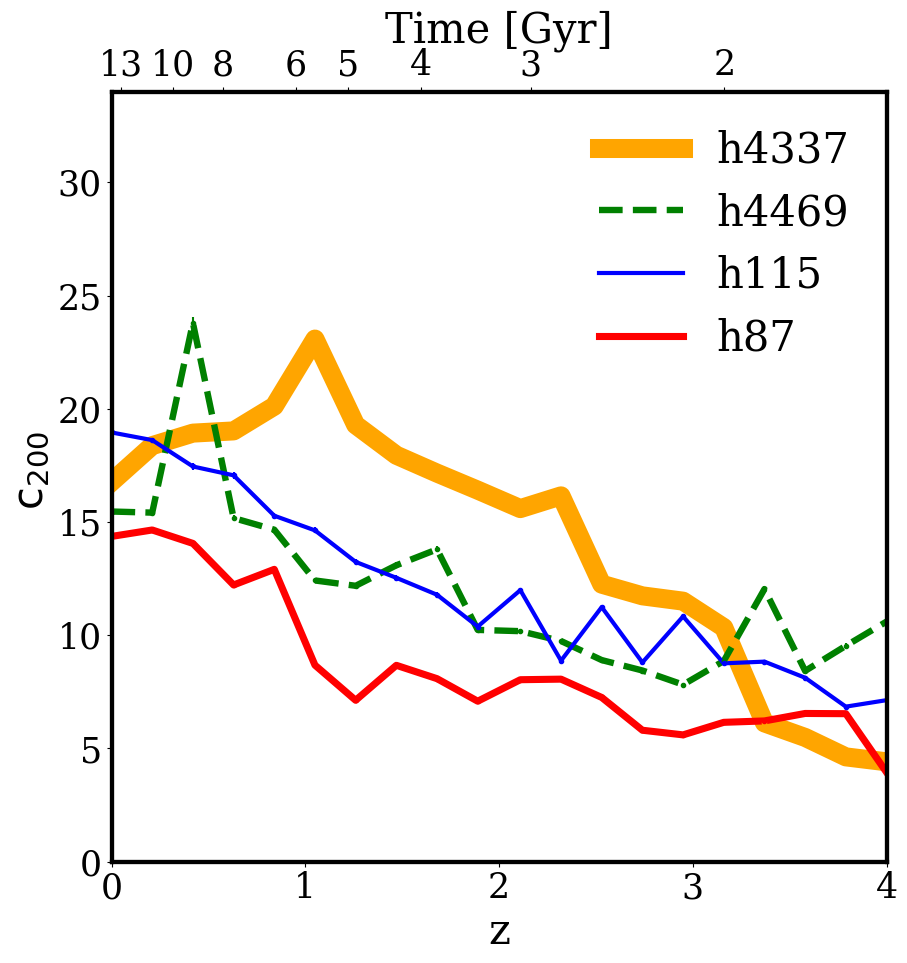}
\includegraphics[width=0.47\columnwidth]{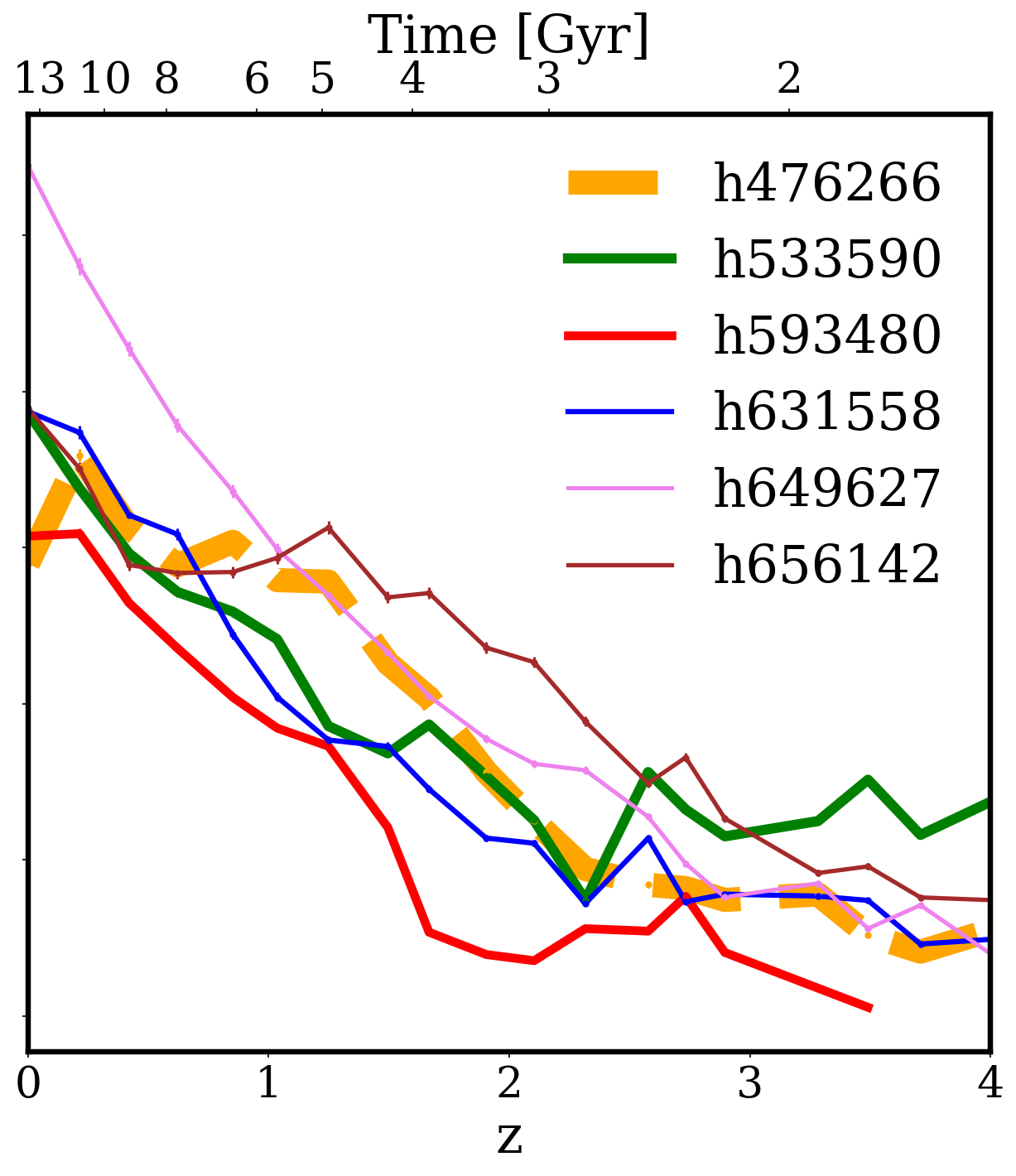}
\caption{Halo concentration as a function of the redshift, $z$, for the \cielo~ (\textit{left panel}) and TNG50 (\textit{right panel}) haloes. The line widths represent the total halo mass $\mathrm{M_{200}^{tot}}$, where thicker lines indicate the most massive systems (see Tab.~\ref{Tab:simu}). Dashed lines indicate the haloes with recent merger. Our results show that the halo concentration increases as the redshift decreases.}
\label{fig:cNFW_z}
\end{figure}

Firstly, we study the DM density profile of the individual haloes selected at different redshifts. We estimate the halo structural parameters by fitting to the mass profile the NFW model \citep{Navarro1997}:

\begin{equation}
    \mathrm{\rho (r)= \frac{\rho _{s}}{\left(\frac{r}{r_{s}}\right)\left(1+\frac{r}{r_{s}}\right)^2}},
\end{equation}

\noindent where $\mathrm{r_{s}}$ and $\mathrm{\rho_{s}}$ are the scale radius and density. In general, the NFW profile provides a good fit to the spherically-averaged $\rho (r)$ profiles. In fact, for all redshift analysed, we estimate the standard error associated with the parameter $\mathrm{r_{s}}$. In both simulations, the fitting errors were below $< 1.6\%$. The best-fit NFW profiles yield estimates of the halo structural parameters $\mathrm{r_{s}}$ and $\mathrm{\rho_{s}}$ for each halo in our sample, which we use in turn to estimate the concentration parameter $\mathrm{c_{200}= r_{200}/r_{s}}$, shown in Fig.~\ref{fig:cNFW_z}. 

Overall, we find that the concentration parameter $\mathrm{c_{200}}$ evolves towards higher values (i.e. more concentrated) in time for both the \cielo~ and TNG50 haloes in agreement with previous works \citep{Gao2008, Ludlow2014}. Fig.~\ref{fig:cNFW_z} also show, in terms of line widths in the plots, the relation of the halo mass with $\mathrm{c_{200}}$. We do not find a clear dependence in both halo selections between halo mass and concentration. \citet{Ludlow2014} found that the concentration is a monotonic but weak function of mass, varying by only a factor of $\sim$ 4 over a mass range of $\mathrm{M_{200}=10^{10}-10^{15}M_{\odot}}$ at $z=0$. Our halo selection only changes one decade in mass, which can explain our results. This seemingly complex mass-redshift-concentration dependence has been described using the dimensionless 'peak height' mass parameter $\mathrm{\nu (M,z)=\delta _{crit}(z)/\sigma(M,z)}$, where $\mathrm{\sigma (M,z)}$ is the linear fluctuation at $z$ in spheres of mass $\mathrm{M_{halo}}$ \citep{Ludlow2014}.

Additionally, the evolution of baryon accretion contributes to contract the inner region of haloes (see Fig.~\ref{fig:MAH_b}, to follow the MAH of baryons). The evolution of halo concentration is better reproduced by models that link the concentration of a halo with its mass accretion history. The concentration is empirically found to trace the time when halos transition from a period of “fast growth” to another where mass is accreted more gradually. \citep{Wechsler2002,Zhao2003,Lu2006}.

\begin{figure}
\centering
\includegraphics[width=0.525\columnwidth]{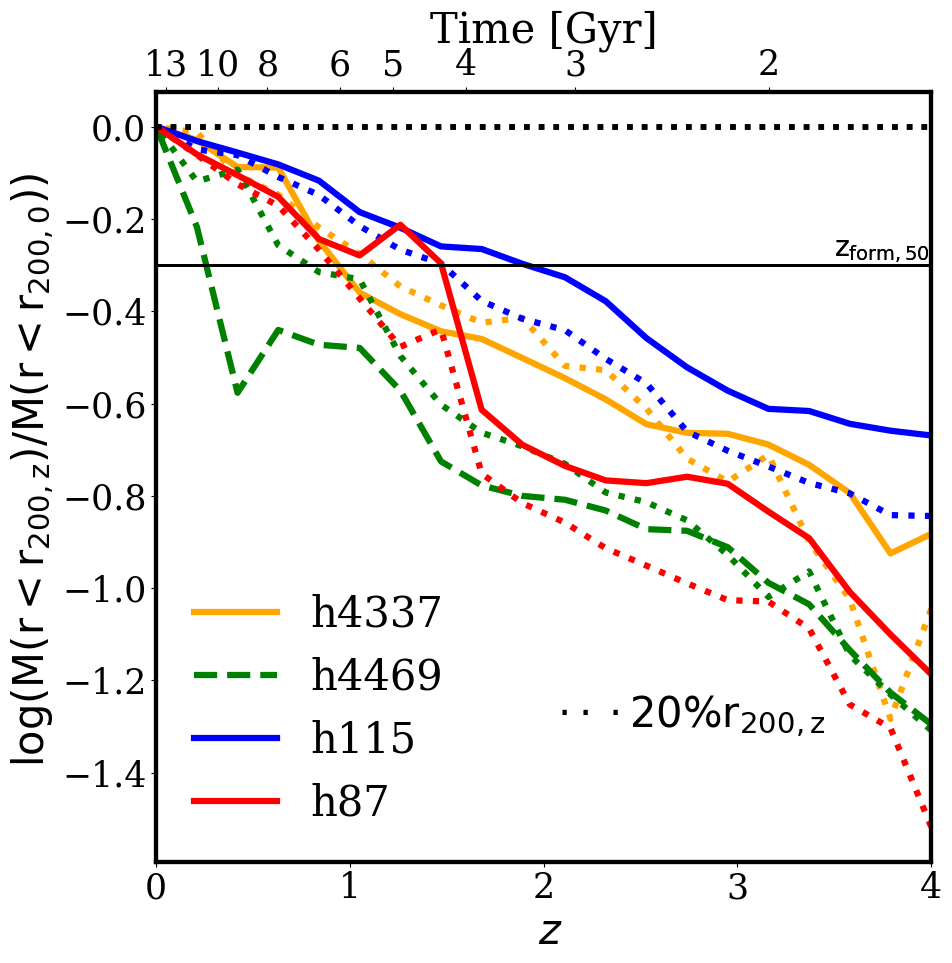}
\includegraphics[width=0.46\columnwidth]{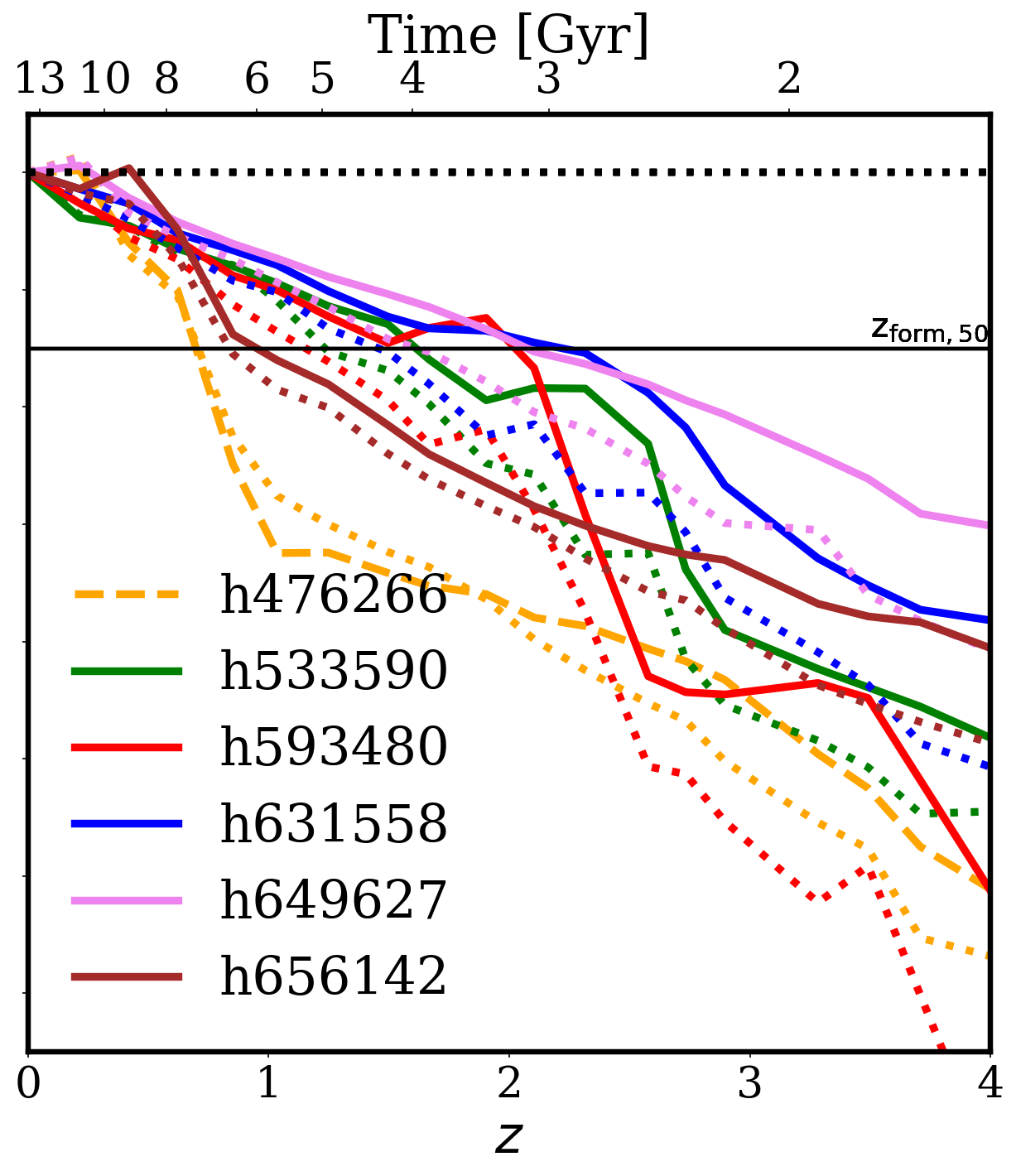}
\caption{The cosmic evolution of the mass accretion history (MAH) normalized by the halo mass at $z=0$ for the DM halos of \cielo{} (\textit{left}) and TNG50 (\textit{right}). Dashed lines indicate the haloes with recent mergers while dotted lines indicate the MAH within $\mathrm{r< 20\%r_{200}}$ for each halo. Black horizontal lines are used as reference to estimate the formation time of the haloes, as the redshift at which the mass of the halo reach half of their mass at $z =0$ ($\mathrm{z_{form,50}}$).}
\label{fig:MAH}
\end{figure}

Motivated by these findings, in Fig.~\ref{fig:MAH} we show the MAH of DM within the virial radius (filled lines) and the correspondent inside the $\mathrm{20 \% r_{200}}$ (dotted lines). The black horizontal lines indicate when the halo reaches half of its final mass at $z=0$ ($\mathrm{z_{form,50}}$). The effects of major accretion due to material stripped from a close satellite (h4469) or due to merger activity (h476266) in recent times appears as sudden peaks in the curve of MAH, a product of a gain or loss in mass. Merger events can also be spot at $\mathrm{20 \% r_{200}}$ with smoother changes with respect to the outer regions of the haloes. There is a general trend of haloes increasing their mass over time as expected, with haloes with lower merger activity in recent times reaching sooner their $\mathrm{z_{form,50}}$ in comparison with the ones with major accretion at $z \sim 0$ (h4469 and h476266). The latter have to gain most of their final mass until late times.  

The increment of mass in recent times can be studied in Fig.~\ref{fig:rate_accretion} and in Tab.~\ref{Tab:time_formation} in the Appendix section, inspecting the instantaneous halo growth at $z \sim 0$ (i.e. $\mathrm{dlog(M)/dt_{z \sim 0}}$) and the formation redshift at $70 \%$ ($\mathrm{z_{form,70}}$). Haloes with merger activity at late time report the higher $\mathrm{dlog(M)/dt_{z \sim 0}}$ and later formation redshifts, $\mathrm{z_{form,70}}$, among the halo selection. 

In agreement with \citet{Zavala2016} and \citet{Lagos2017} we observe two different regimes in the evolution of the virial radius $\mathrm{r_{200}}$ as we can see in Fig.~\ref{fig:r_vs_z} in the Appendix section. First, a slow increase of halo size over time, up to $z=2$ followed by an acceleration in the increasing of $\mathrm{r_{200}}$, showing the turnaround point. Before that point in time, the DM halo gain angular momentum through tidal torques from their environment until maximum expansion (turnaround point), and afterward the halo collapse into virialized structures that conserve their angular momentum \citep{Doroshkevich1970,White1984,Theuns1996a,Theuns1996b}. 

\subsection{Halo morphology evolution}
\label{sec:Morph}

In \citet{Cataldi2020} we find that for \textsc{EAGLE} and Fenix haloes, baryons have a significant impact on the shape of the inner halo, mainly within $\sim$ 20 percent of the virial radius. In order to dig into the evolution of halo shape, we compute the shapes of our selected halo sample and focus the analysis mainly at $\mathrm{20\% \, r_{200}}$.

We describe them using the semi-axes of the triaxial ellipsoids, $\rm a>b>c$, where $\rm a$, $\rm b$ and $\rm c$ are the major, intermediate and minor axis respectively of the shape tensor $\rm S_{ij}$ \citep[e.g.][]{Bailin2005, Zemp2011}. Here we use an iterative method that starts with particles within a spherical shell \citep[i.e. $ \rm q=s=1$][]{Dubinski1991, Curir1993}.  

\begin{figure*}
\centering
\includegraphics[width=0.7\textwidth]{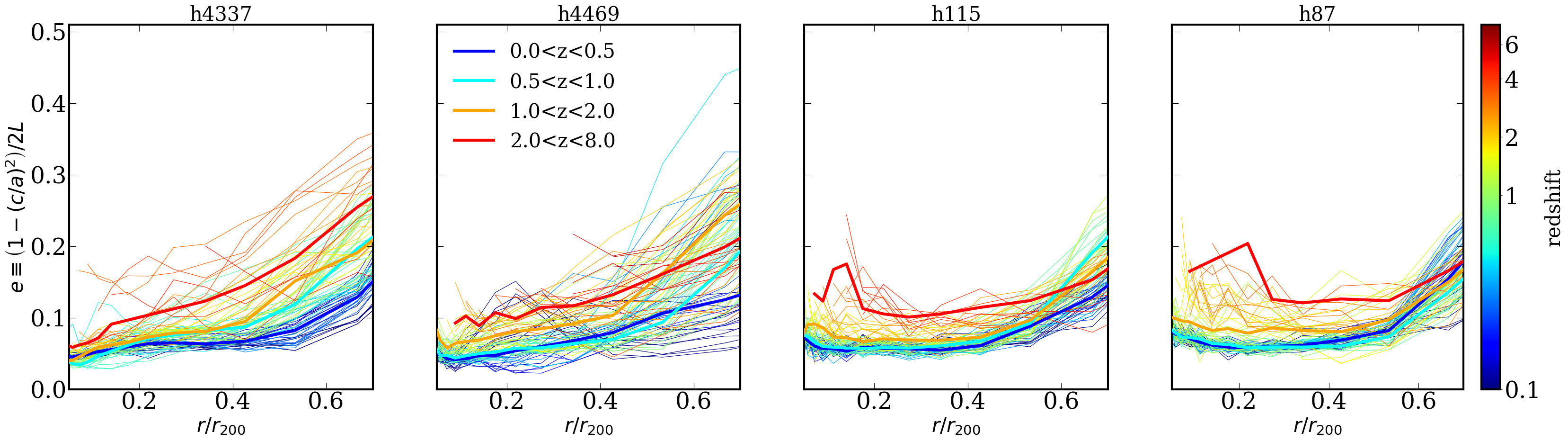}
\includegraphics[width=\textwidth]{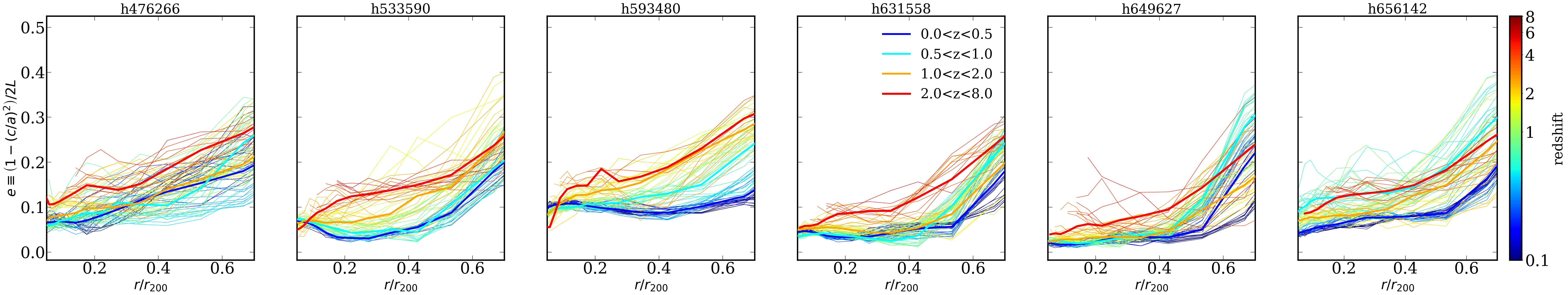}
\caption{The evolution of the shape parameters as a function of the distance to the halo center. Each panel represents a different halo, as indicated by the label. We show the ellipticity $\mathrm{e \equiv \left ( 1 - (c/a)^{2} \right )/2L}$ for the \cielo{} (\textit{top panels}) and TNG50 (\textit{bottom panels}) haloes versus $\mathrm{r/r_{200}}$, colored by the redshift between $0<z<8$. We also show the median values for four subsample in redshift bins. Blue lines for $0<z<0.5$, cyan for $0.5<z<1.0$, orange $1.0<z<2.0$ and red lines $2.0<z<8.0$, according to the redshift color coded. In all cases for our selected haloes, the ellipticity increase for outer radii and for higher redshifts.}
\label{fig:e_r_all}
\end{figure*}

To obtain the ratios $\rm q\equiv b/a$ and $\rm s\equiv c/a$, we diagonalise the reduced inertia tensor to compute the eigenvectors and eigenvalues, as in  \citet{Tissera1998}. Traditionally the $\rm s$ shape parameter has been used as a measure of halo sphericity \citep[e.g.][]{Allgood2005,Vera-Ciro2014, Chua2019}. We adopted the triaxiality parameter, defined as $\rm T \equiv (1-q^{2})/(1-s^{2})$, which quantifies the degree of prolatness or oblatness: $\rm T=1$ describes a completely prolate halo ($\rm a > b\approx c$) while $\rm T = 0$ describes a completely oblate halo ($\rm a\approx  \rm b > c$). Haloes with $\rm T > 0.67$ are considered prolate and haloes with $\rm T < 0.33$ oblates, while those with $\rm 0.33 <T <0.67$ are considered triaxial \citep{Allgood2005,Artale2018}. We define the ellipticity as,

\begin{equation}
    \mathrm{e \equiv \left (1-(c/a)^{2} \right )/2L},
\end{equation}

\noindent where $\mathrm{L \equiv 1+(b/a)^{2}+(c/a)^{2}}$. The condition $\mathrm{a \geq  b \geq c}$ implies that the domain of e is the range $\mathrm{[0, 1/2]}$.

In Fig.~\ref{fig:e_r_all}, we show the evolution of the ellipticity as a function of the distance to the halo center. To inspect the evolution in further detail, we extend the range of redshift up to $0<z<8$. For more recent redshifts, the haloes become less elliptical and in correspondence, more spherical. For earlier times, haloes present triaxial shapes, while as they evolve in time their ellipticity  decreases. This result is in agreement with previous works \citep{Allgood2005,Chua2019,Cataldi2020}. Interestingly, the morphologies of \cielo{} haloes present a more ordered trend in their decrease of ellipticity with redshift in comparison with TNG50 haloes, which report a weaker trend in their decrease of ellipticity.

\begin{figure}
\centering
\includegraphics[width=\columnwidth]{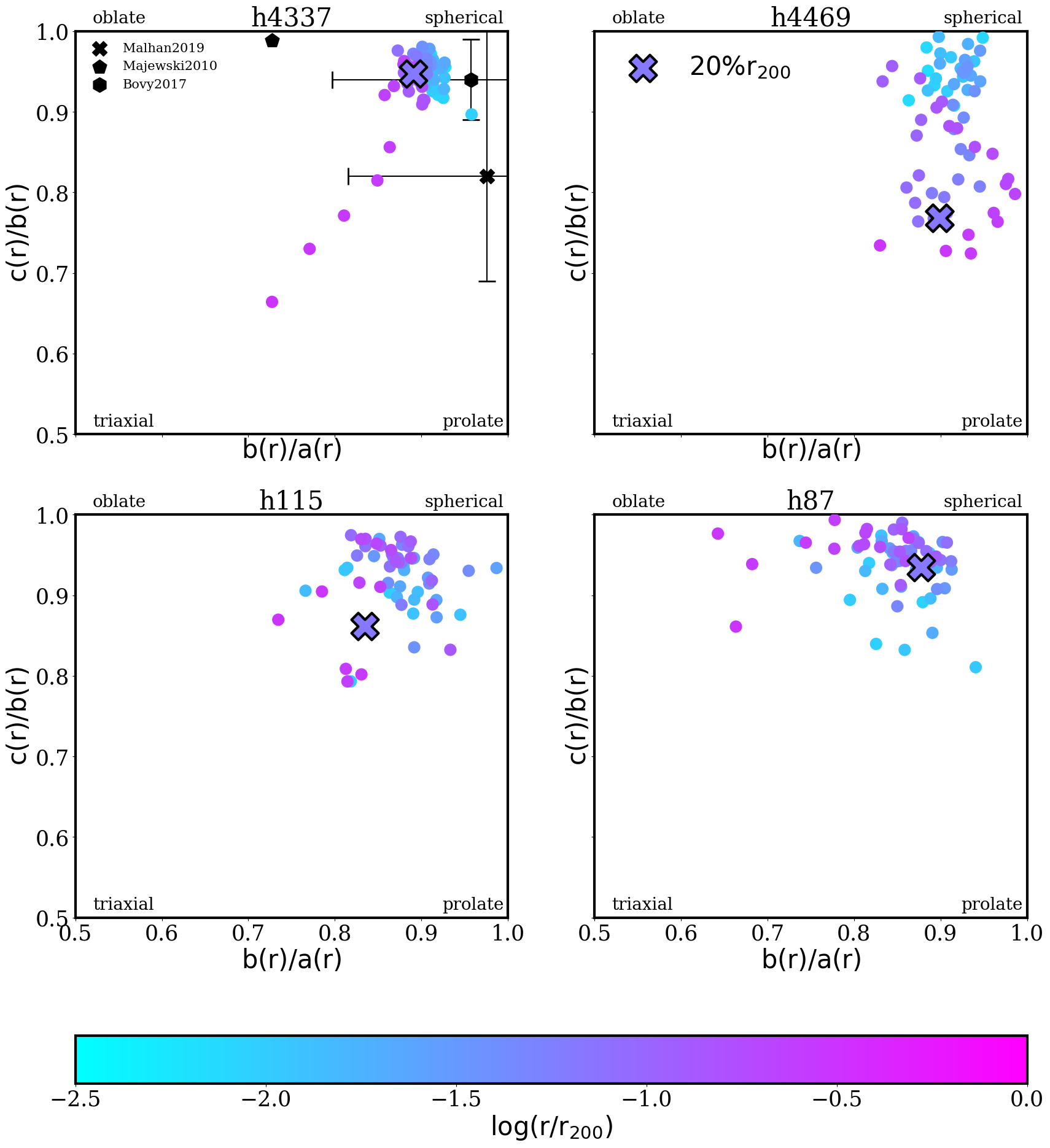}
\caption{The distribution of \cielo{} haloes as a function of their inner halo axial ratios measured for $\mathrm{-2.5<log(r/r_{200})<0}$ at $z=0$. When $\mathrm{b/a \sim 1.0}$ and $\mathrm{c/b \sim 1.0}$, the haloes become more spherical (upper right corner). The regions of parameter space corresponding to spherical, prolate, triaxial and oblate haloes are indicated in each panel. For all the haloes the morphology tends to be more spherical for the inner regions and more triaxial for the outer radii. We indicate the shape ratios at $\mathrm{20\% \, r_{200}}$ with a cross symbol. We compare our results with observational constraints for the DM halo shape in the Milky Way by \citet{Law2010}, \citet{Bovy2016} and \citet{Malhan2019} (see legend and symbols on the top left panel).}
\label{fig:Trayford_radius}
\end{figure}

\begin{figure}
\centering
\includegraphics[width=\columnwidth]{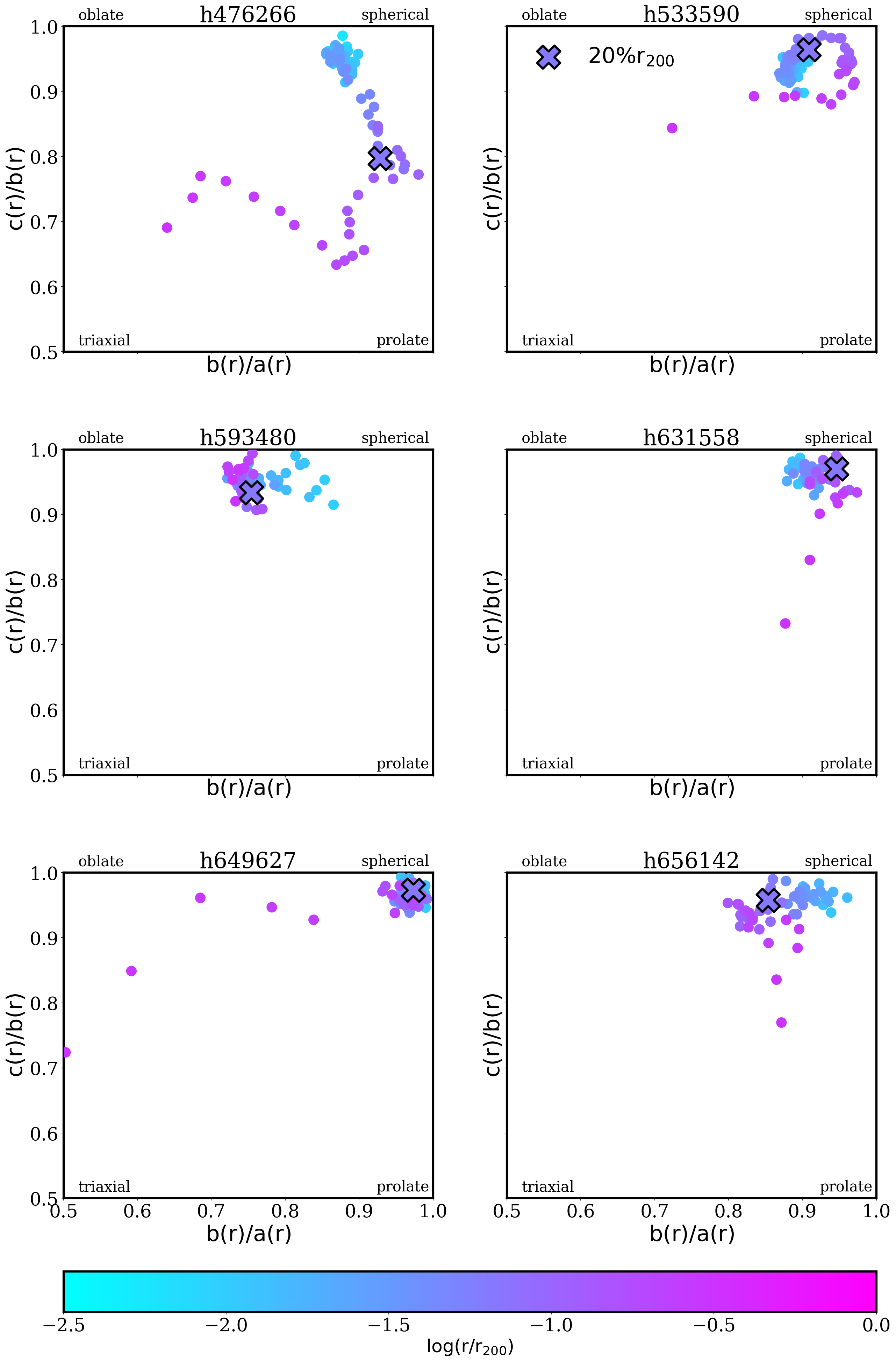}
\caption{Same as Fig.~\ref{fig:Trayford_radius} but for the TNG50 selected haloes at $z=0$. We find that the general trend of the haloes shapes are distributed between spherical and oblate shapes.}
\label{fig:Trayford_radius_TNG}
\end{figure}

A useful way to study the shape evolution of the DM haloes with time and radii is the plane proposed by \citet{Trayford2019}, also applied in previous studies \citep[see,][]{Cataldi2020,Cataldi2022}. However, we modify it to visualize the evolution of the halo axis ratios at different radii. Fig.~\ref{fig:Trayford_radius} and Fig.~\ref{fig:Trayford_radius_TNG} present the results for the \cielo{} and TNG50 haloes samples, respectively, at $z=0$. In our representation, the upper right corner, $\mathrm{b/a \sim  1.0}$ and $\mathrm{c/b \sim 1.0}$, correspond to spherical haloes. The spherical, prolate, triaxial, and oblate regions are labeled correspondingly, and the color code indicates the distance to the halo center. 

For \cielo{}, all haloes morphology tends to be more triaxial for the outer radii and more spherical in the inner regions. In the case of TNG50, haloes in the central regions are distributed between spherical and oblate (h593480) shapes. However, the evolution path to this shape configurations is remarkably different in each case. \citet{Chua2022} studied the halo morphology dependence on radii and for different mass resolution in the TNG50 suite. These authors reported spherical and oblate shapes for haloes at central radii and a small tendency for haloes of TNG50 to be more spherical and oblate than the lower mass resolution simulations of the suite.

The configuration goes from typically spherical in the inner regions (related to the collapse of matter in earlier times) to a triaxial shape in the outskirts (corresponding to the shells that have collapsed more recently through a preferred direction). As the haloes evolve with time, there is also the effect of baryonic condensation to form the galaxy at the inner region of the halo (see Fig.~\ref{fig:MAH_b}). This baryonic concentration contributes to rounding up the halo morphology. 

The halo shape structure at earlier times is imprinted in the present day ($z=0$) shape trends with radius \citep{Vera-Ciro2012}. Therefore, the different paths to sphericity can be studied analyzing the evolution of halo morphologies. In Fig.~\ref{fig:Trayford_radius} and Fig.~\ref{fig:Trayford_radius_TNG} we also show the halo shape at $\mathrm{20\%  \, r_{200}}$ with cross symbols. We compare the shapes at the galactocentric radii, where the stellar disc is located, which is the radius that maximize the effects of baryonic concentration \citep{Cataldi2020}. 

\begin{figure}
\centering
\includegraphics[width=\columnwidth]{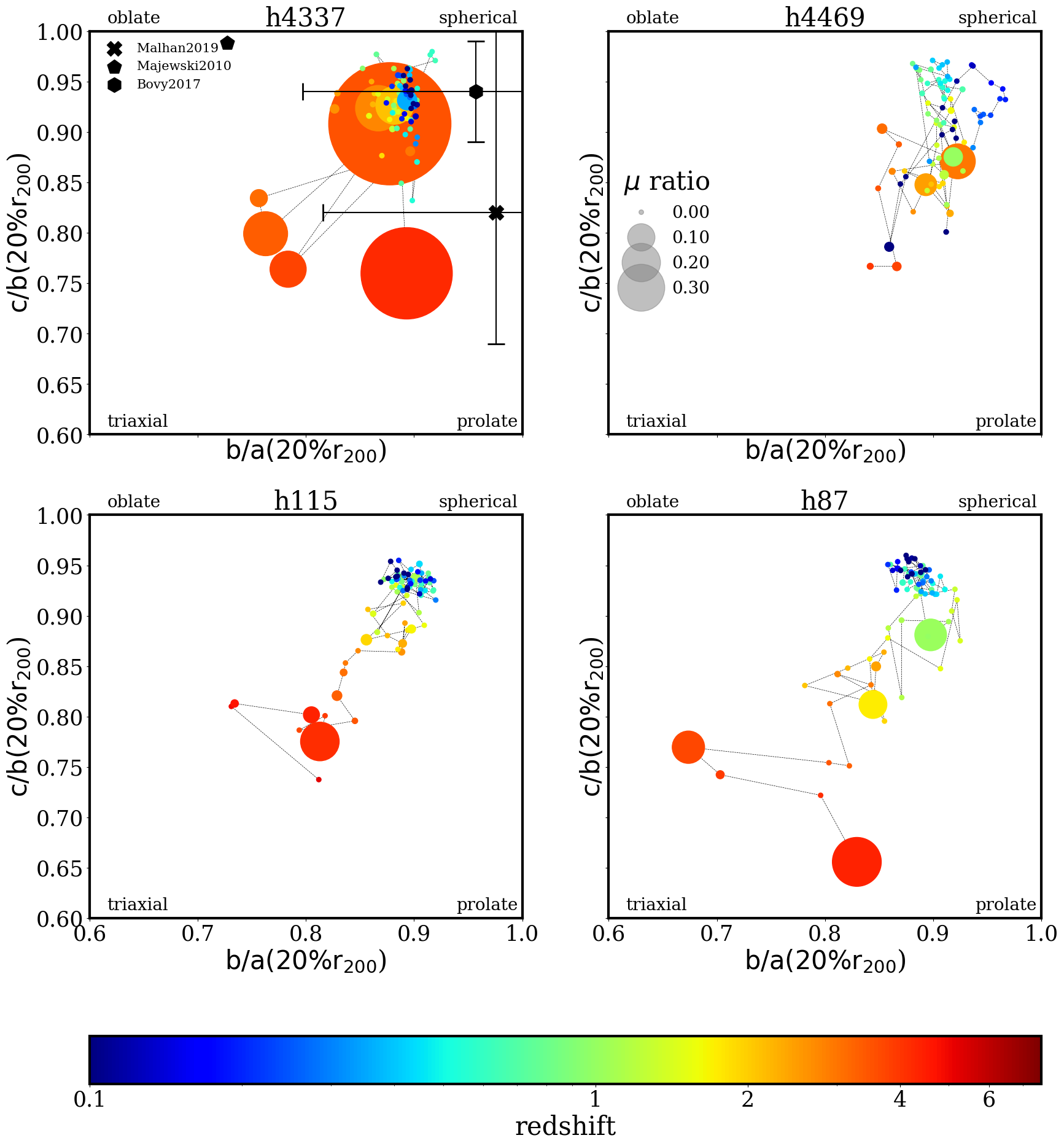}
\caption{Same plane as Fig.~\ref{fig:Trayford_radius} but for the inner axial ratio $\mathrm{20\% \, r_{200}}$ at different redshifts, for the \cielo{} haloes. The size of the circles is proportional to the merger stellar ratio, $\mu$. We find that haloes tend to be more spherical at lower redshift. Inspecting the merge rate, for a greater merger event in a given redshift (marked with bigger symbols) bigger changes in halo morphology. In this figure, we extended the redshift range up to $0<z<8$. Black symbols on the top left panel are the observational constraints for the DM halo shape in the Milky Way by \citet{Law2010}, \citet{Bovy2016}, and \citet{Malhan2019}. The overall evolution across the shape parameter space is for haloes evolution from triaxial to spherical configuration. The effect of recent major accretion can be observed in h4469, with a backward tendency to be less spherical at $z \sim 0$.}
\label{fig:Trayford_LG}
\end{figure}

\begin{figure}
\centering
\includegraphics[width=\columnwidth]{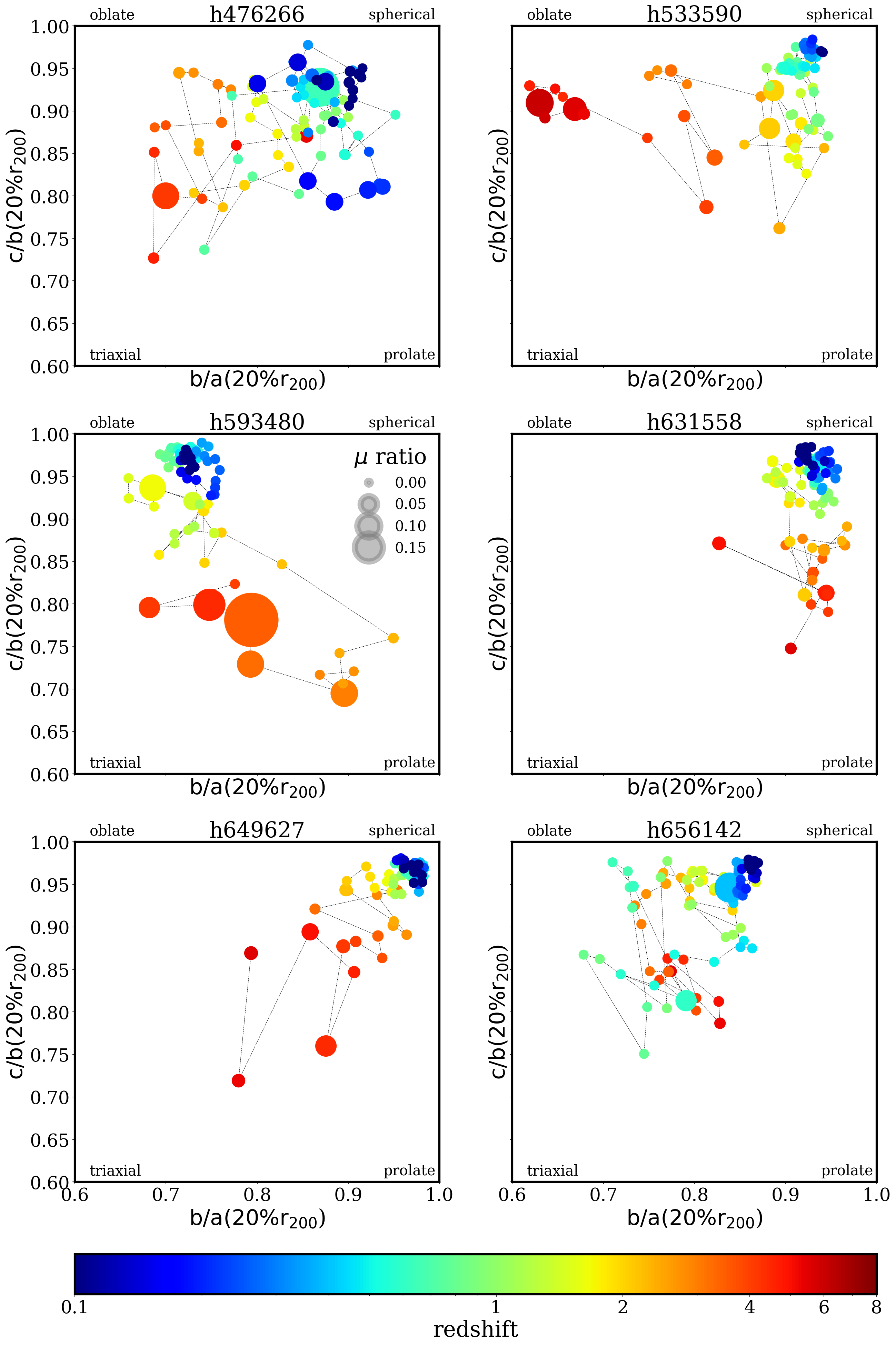}
\caption{Same as Fig.~\ref{fig:Trayford_LG} but for the TNG50 selected haloes. In contrast to \cielo{} haloes, not all TNG50 haloes end up in a spherical configuration. Interestingly, the halo h593480 stabilizes its configuration with an oblate morphology, and the halo h476266 (with recent mergers activity) with a final spherical shape but with a great dispersion.}
\label{fig:Trayford_TNG}
\end{figure}

To shed light on this, we analyse the correlation between merger events and the resulting morphology in Fig.~\ref{fig:Trayford_LG} and Fig.~\ref{fig:Trayford_TNG}. We show the axial ratios $\mathrm{c/b}$ vs. $\mathrm{b/a}$ within $\mathrm{20\%  \, r_{200}}$. Here each point is colored according to the redshift. The size of the dots is proportional to the ratio $\mu$, a quantitative measure of the merger events at a given redshift. 

The morphology of DM haloes, although varies greatly, presents a trend in the sense of haloes being more spherical in more recent times. These trends are clearer in \cielo{} than TNG50 samples. The effects of recent major accretion (h4469) and recent merger activity (h47626) were found with larger dispersion reaching lower redshifts, although with a path to more spherical shapes.

\begin{figure}
\centering
\includegraphics[width=\columnwidth]{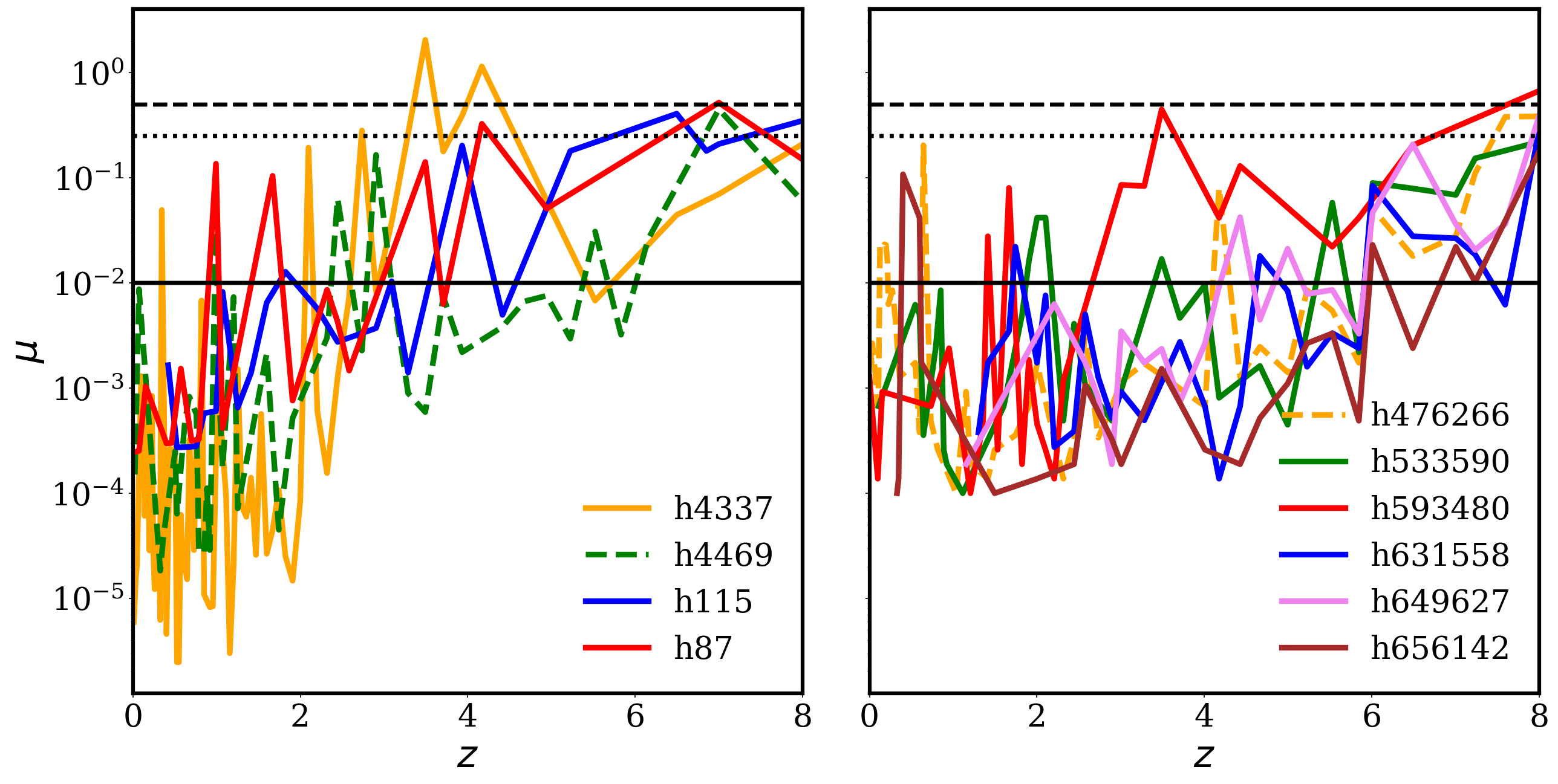}
\caption{The evolution of the merger stellar ratio, $\mu$ for the \cielo{} (\textit{left panel}) and TNG50 haloes (\textit{right panel}). Dashed lines indicate the haloes with no constraints in mergers activity. Black horizontal lines correspond to $\mu = 0.01$, while dotted and dashed lines correspond to $\mu = 0.25$ and $\mu = 0.50$, respectively.}
\label{fig:mu_vs_z}
\end{figure}

Fig.~\ref{fig:mu_vs_z} shows the evolution of the merger stellar ratio, $\mu$, across cosmic time for all selected haloes in each simulation. The last major events ($\mu > 0.25$) occurred at $z \sim 3$ in the case of h4437 and at $z \sim 6$ for the rest of \cielo{} haloes. In the case of TNG50, the last major event is reported at $z \sim 4$ for h593480 and at $z \sim 7$ for the rest of the selection.

The differences in merger histories between \cielo{} and TNG50 could be explained due to the different environments for halo selection. \cielo{} haloes were chosen to correspond to an analogue environment of the Local Group. In the case of TNG50 haloes, there were no environmental constraints. The difference in where the haloes were embedded in the cosmic web, leads to a higher merger activity for the TNG50 at late times, which results in a much wider variety of morphologies configuration at $z \sim 0$.
 
Although being evident that recent major merger affects DM shapes, the exact connection is not yet fully understood. However, it is clear that the structure of individual haloes is closely related to their merger history. Shape changes, for instance, have been linked to the properties of the last major merger \citep[e.g.][]{Despali2017} and the remnant has been found to be elongated along the merger axis \citep[e.g.][]{Maccio2007,Vera-Ciro2012}.

We also find that haloes with greater changes in morphology through redshift are correlated with the number and importance of the merger events. This effect is even more significant for h476266 in TNG50 with recent merger activity. Mergers can be followed by slow accretion along filaments until the cluster ends up in a relatively viralized final phase with a nearly regular and spherical shape. For recent redshifts (see blue dots Fig.~\ref{fig:Trayford_LG} and Fig.~\ref{fig:Trayford_TNG}), the merger activity weakens for both simulations and in consequence the relaxation times increase, contributing to more spherical shapes.

\subsection{Effects of mass infall}
\label{sec:Infall}

In the previous section, we discussed two main physical processes that contribute to halo shape evolution: the condensation of baryons in the inner regions and the infall of matter in the outer shells of the halo. While the presence of baryons tends to round up the halo, the infall through filaments produces differences in axis length and, in consequence, contributes to more elongated ones \citep{Despali2014, Gouin2021}. In this section, we discuss the influence of the cosmic web environment on the DM halo shapes. 

Using N-body simulations, previous studies investigate the correlation between the environment and shape of haloes \citep[e.g.][]{Libeskind2011,Vera-Ciro2012}. DM haloes grow over time fed by the surrounding density field, through continuous injection of matter. This accretion may be secular or in a series of more violent mergers. 

Through the virialization process, each halo acquires a new equilibrium configuration as new material is accreted into the gravitational potential well. The preferential direction of the infalling material in the cosmic environment is usually given by the filaments, whereas a more isotropic mode is expected when the halo is embedded in a large structure \citep[e.g][]{Wang2011,Vera-Ciro2012,Shao2021,Baptista2022}. 

\begin{figure*}
\centering
\includegraphics[width=\textwidth]{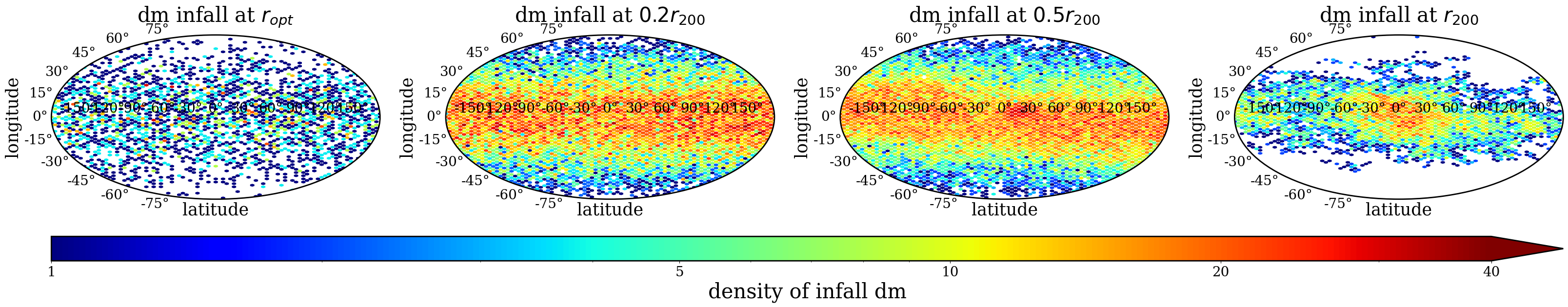}
\includegraphics[width=\textwidth]{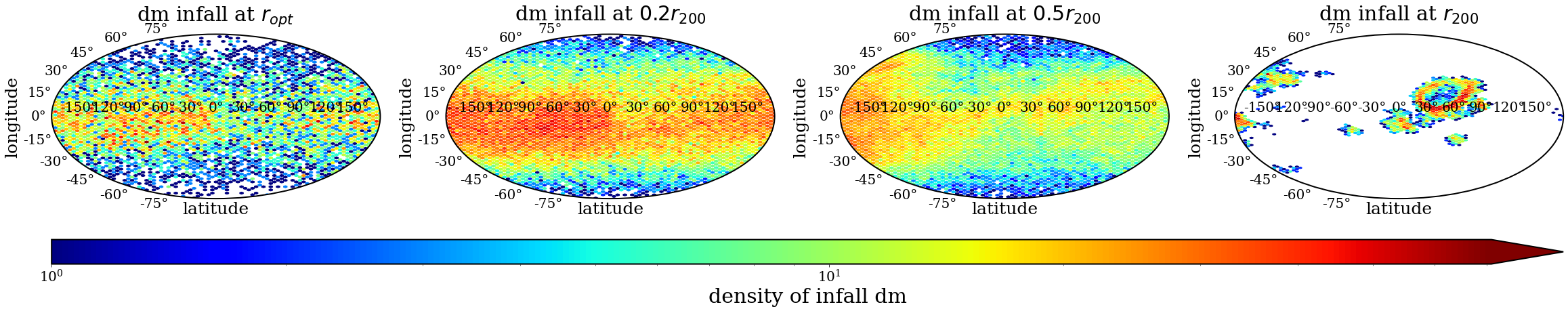}
\caption{The projected infall DM particles for the halo h4337 (\textit{top panels}) and h593480 (\textit{bottom panels}). From left to right, each panel is computed on concentric shells of radius $\mathrm{r_{opt}}$, $\mathrm{20 \% r_{200}}$, $\mathrm{50 \% r_{200}}$ and $\mathrm{r_{200}}$. Each time a DM particle is accreted across one of these shells, its entry point is recorded and plotted. The red regions indicate high density while the blue ones indicate low density.}
\label{fig:Aitoff}
\end{figure*}

As an example of the aforementioned, we show in Fig.~\ref{fig:Aitoff} the Aitoff map of the projected DM particles infall at $z=0$ for two haloes, h4377 from \cielo{} (top panel) and h593480 from TNG50 (bottom panel). The map projected is centered around the stellar disc frame. At each output time, we select particles with negative radial velocity pointing towards the center of mass of a halo, $\mathrm{v_{r}<0}$ (infalling particle), in different spherical shells: $\mathrm{1.0<r/r_{opt}<1.2}$ \footnote{The optical radius, $\rm r_{\rm opt}$, is define as the radius that encloses 80 percent of the baryonic mass (gas and stars) of the galaxy.},  $\mathrm{0.15<r/r_{200}<0.25}$, $\mathrm{0.45<r/r_{200}<0.55}$ and $\mathrm{1.0<r/r_{200}<1.2}$.  

\begin{figure}
\centering
\includegraphics[width=\columnwidth]{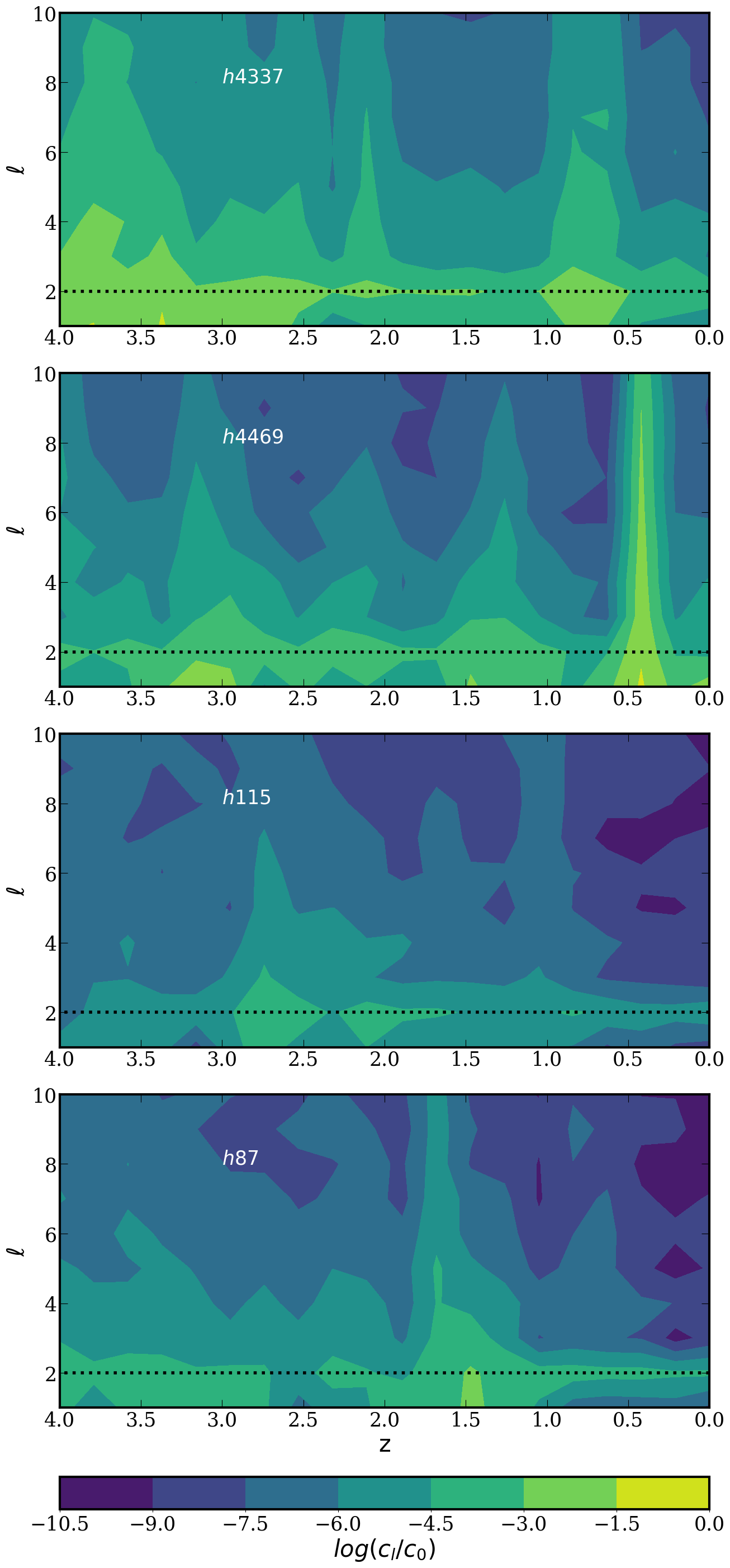}
\caption{Multipole expansion of the infalling DM particles ($\mathrm{v_{r}<0}$) in the region $\mathrm{1.0< r/r_{200} < 1.2}$ as a function of redshift for the haloes in \cielo{} simulation. The color map indicates the ratio $\mathrm{log(C_{\ell}/C_{0}}$). After the monopole ($\ell = 0$), the quadrupole mode ($\ell = 2$) is the main component of the DM particle's infall suggesting that filaments are a fundamental ingredient to understanding the mass accretion history and halo shape.}
\label{fig:Map_CIELO}
\end{figure}

\begin{figure}
\centering
\includegraphics[width=\columnwidth]{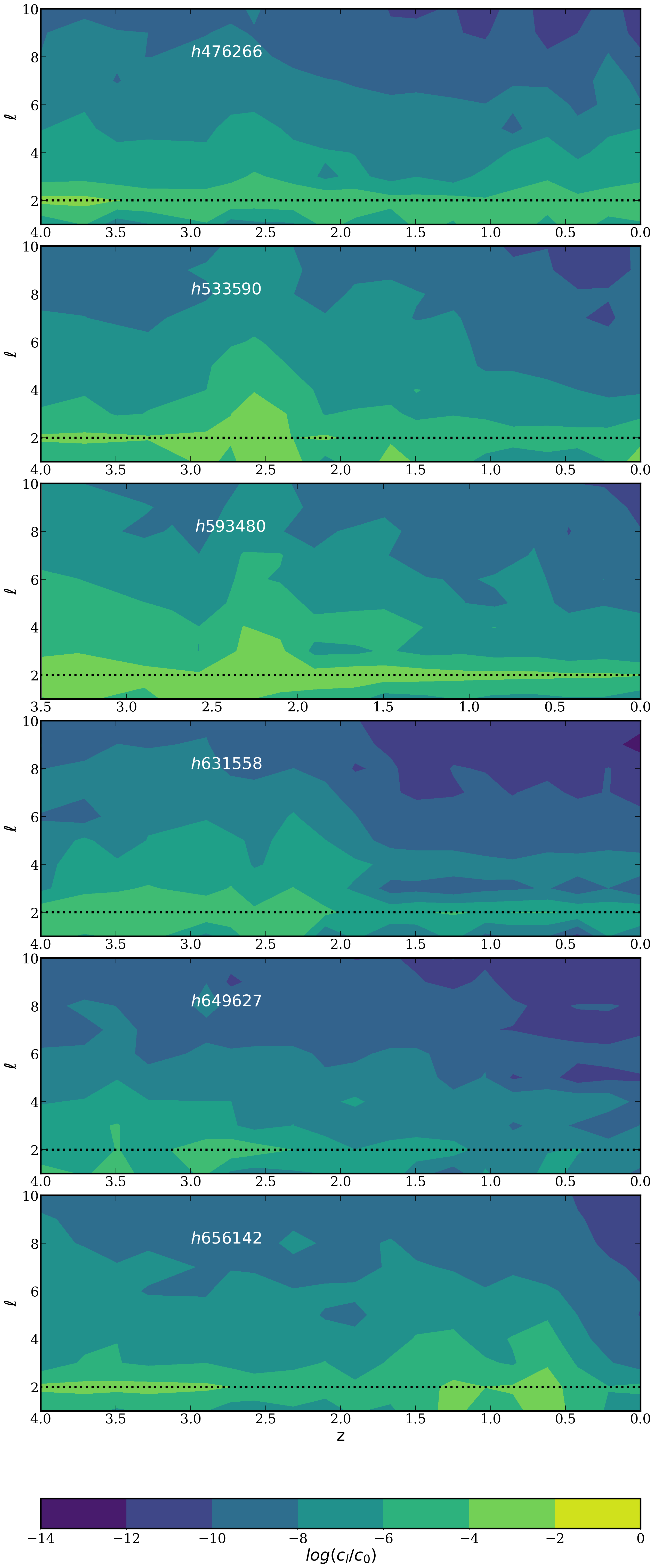}
\caption{Same as Fig.~\ref{fig:Map_CIELO} for the TNG50 haloes. TNG50 haloes had a deeper contribution to filament accretion than \cielo{} haloes.}
\label{fig:Map_TNG}
\end{figure}

Regions, where the infalling material is larger at a given redshift, correspond to regions with a major overdensity \citep[e.g.,][]{Libeskind2011}. As a result of this, the density retains the configuration of where the mass was accreted. Albeit weaker with decreasing radii, the self-similar patterns across the four shells are present, as the accreted material goes from outer radii to more central regions. 

This different form of material infall to the haloes structure follows a specific distribution on the sky. For instance, whereas isotropic accretion would indicate a uniform signal in the sky, a bi-modal distribution of points in two opposite directions would lead to the presence of a thin filament \citep{Tormen1997,Colberg1999, Libeskind2011,Vera-Ciro2012}. A multipole expansion of the infalling particles in the sky at a given time could be quantified through the power spectrum for the mode $\ell$: 

\begin{equation}
    \mathrm{C_{\ell}=\frac{1}{2\ell+1}\sum _{m=-\ell}^{\ell}\left | a_{\ell}^{m} \right |^2},
\end{equation}

\noindent where the expansion coefficients are,

\begin{equation}
    \mathrm{a_{m}^{l}= \frac{1}{N}\sum_{i=1}^{N}Y_{\ell}^{m}(\theta _i,\phi _i)},
\end{equation}

\noindent where the subscript $i$ indicates the $i$-th particle crossing the chosen shell in the specific angular position, with a negative radial velocity ($\mathrm{v_{r}<0}$). The number of particles, N, can be represented with $\mathrm{N= 4m / (\pi r_{200}^{2})}$.

The $\ell = 0$ term is the monopole, representing in this scheme the isotropic accretion. The $\ell = 2$ corresponds to the quadrupolar moment, meaning that the accretion occurs through a well-defined direction in the space. Similarly, accretion corresponding to more than one preferential direction will shift the power towards higher moments. When a satellite occupies a large area of the sky the configuration will then resemble a dipole and the power spectrum will exhibit higher power in the $\ell = 1$ mode \citep[][]{Vera-Ciro2012}.

We describe the infall of DM particles and the self-similar distribution across a large radial extent. We take the outskirts of the chosen shells (at $\mathrm{r=r_{200}}$) and compute the corresponding amplitude ($C_{\ell}$) of the spherical distribution. For a given DM infall particle with coordinates, $\theta _{i}$ (longitude) and $\phi _{i}$ (latitude), we evaluate the spherical harmonic function for a given $\ell$. The distribution of the infall particles can be approximated by a smooth angular surface density constructed by summing over all $m$ and $\ell$ up to $\ell _{max}$ according to the following equation 

\begin{equation}
    \mathrm{\sigma(\theta,\phi) =\sum _{\ell =0}^{\ell_{max}}\sum_{m=-\ell}^{\ell}a_{\ell}^{m}Y_{\ell}^{m}(\theta _i,\phi _i)}
\end{equation} 

In Fig.~\ref{fig:Map_CIELO} and Fig.~\ref{fig:Map_TNG} we show the multipole expansion of the infalling material ($\mathrm{v_{r}<0}$) in the region $\mathrm{1.0<r/r_{200}<1.2}$ as a function of redshift for \cielo{} and TNG50 haloes. The color map are coded in terms of $\mathrm{log(C_{\ell}/C_{0}}$). In all haloes, the monopole mode is the dominant configuration of infalling material, followed by the quadrupole ($\ell=2$, i.e filaments). The strength of the filament varies across haloes and redshift. 
Compared to the smooth accretion, we expect that satellite infall events to excite a wide range of modes with similar power. Major accretion such as that of halo h4469 at $z \sim 0.4$ is reflected by a rapid peak excitement across all modes (see also Figure~\ref{fig:mu_vs_z}).

\begin{figure}
\centering
\includegraphics[width=\columnwidth]{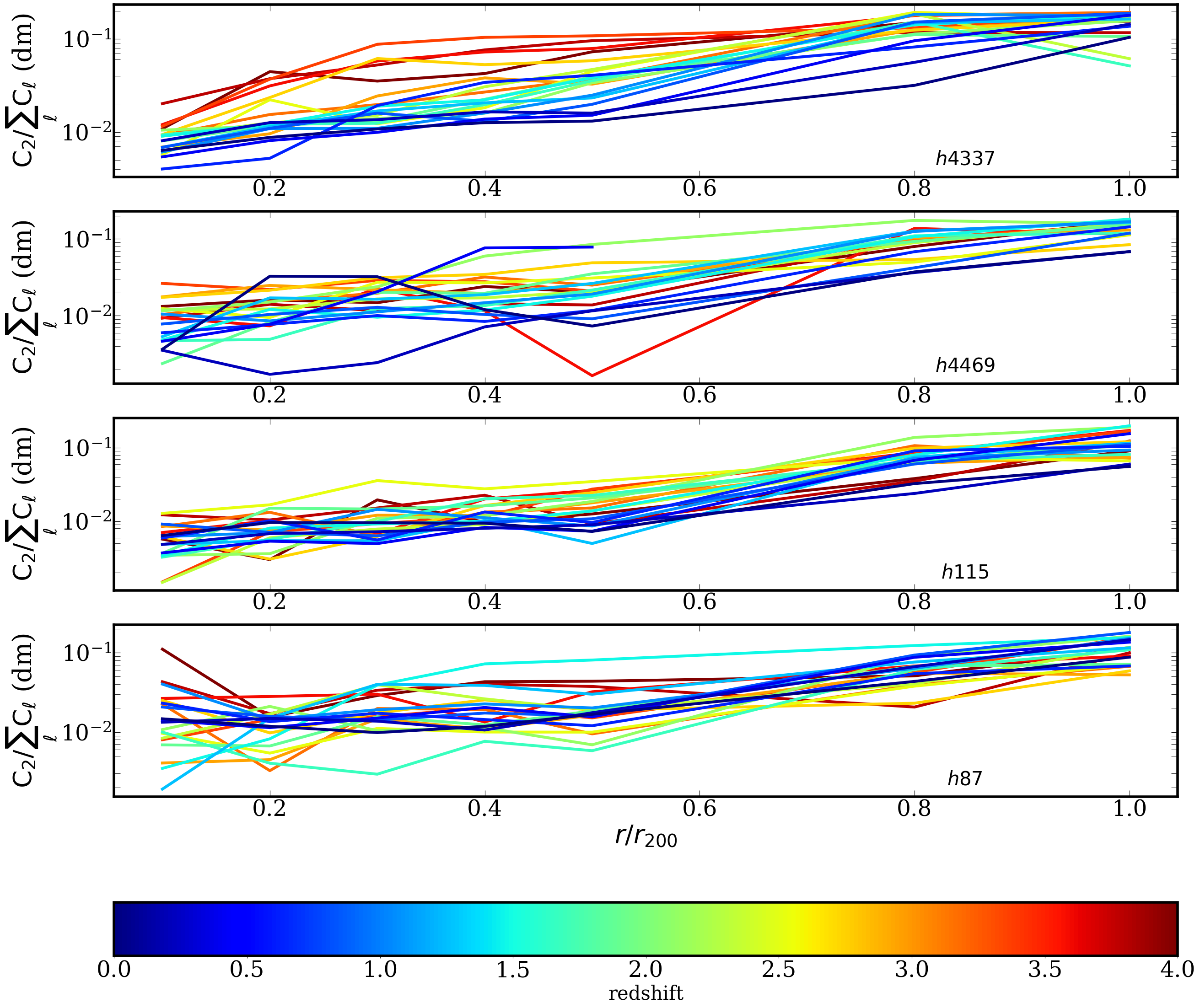}
\caption{Relative contribution of the $\ell=2$ mode to the total power spectrum as a function of $\mathrm{r/r{200}}$ for the haloes in the \cielo{} simulation at different redshifts. $\mathrm{C_{2}/\sum _{\ell}C_{\ell}}$ provides information about the material infalling along a filament. The strength of quadrupole, a privileged direction of infalling mass, decreases with radii as in central regions the motion of the particles is dominated by other physical processes. At half the virial radius ($\mathrm{\sim 50 \% r_{200}}$) we find a trend for the quadrupole strength to decrease steadily with time. This trends weakens for other regions.}
\label{fig:ratiomodes}
\end{figure}

\begin{figure}
\centering
\includegraphics[width=\columnwidth]{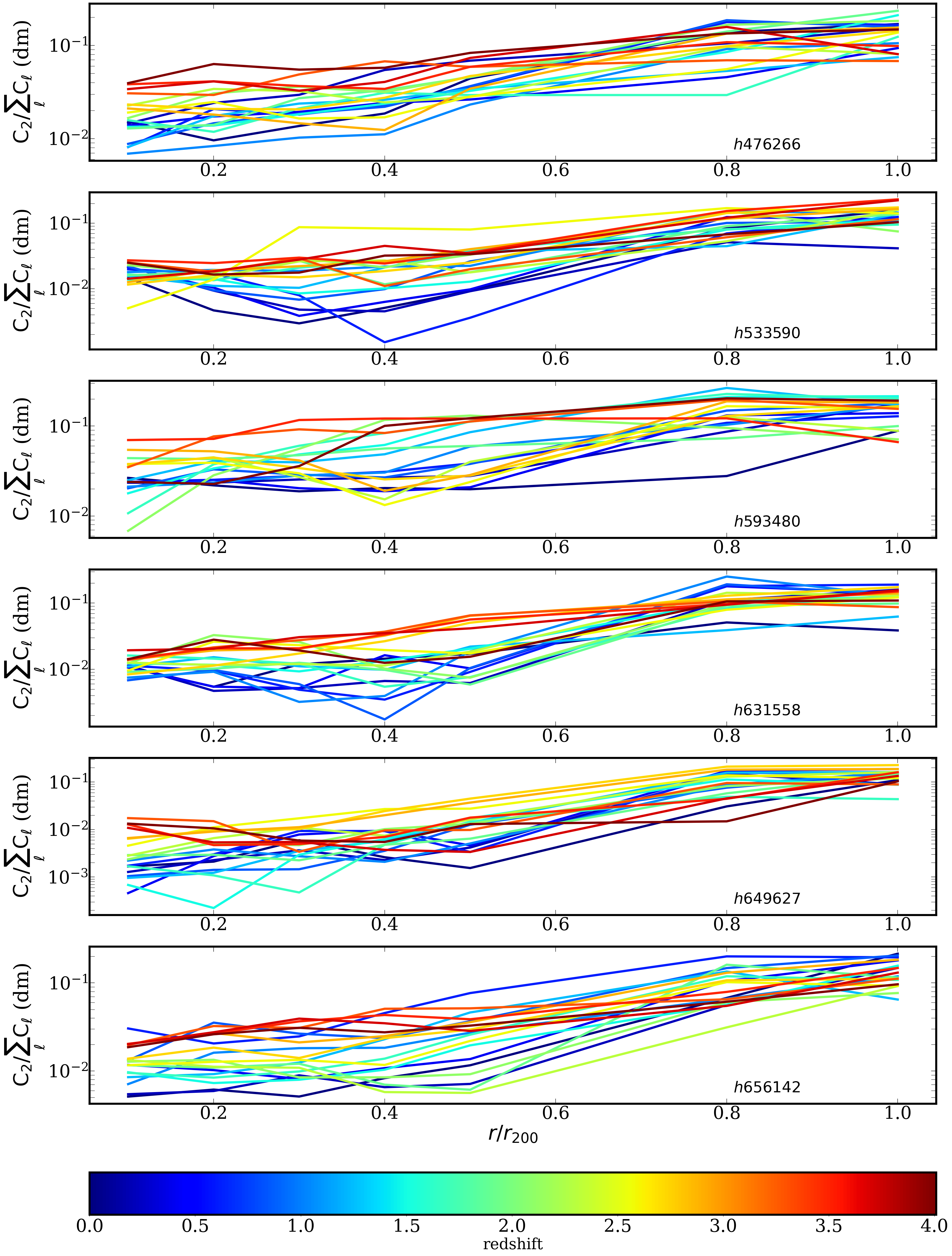}
\caption{Same as Fig.~\ref{fig:ratiomodes} but for the TNG50 selected haloes.}
\label{fig:ratiomodesTNG}
\end{figure}

Fig.~\ref{fig:ratiomodes} and Fig.~\ref{fig:ratiomodesTNG} show the evolution of filament strength by quantifying the relative contribution of the $\ell=2$ mode to the total power spectrum  ($\mathrm{C_{2}/\sum _{\ell}C_{\ell}}$) as a function of $\mathrm{r/r_{200}}$. In each panel, the lines represent different redshifts. The quadrupole strength decrease with radii. In central regions, the motion of the DM particles is expected to be dominated by other velocities configurations \citep[e.g. tube orbits][]{Zhu2017}. Additionally, in Fig.~\ref{fig:ratiomodes} and Fig.~\ref{fig:ratiomodesTNG}, at $\mathrm{r \sim 50\% r_{200}}$, the quadrupole strength decreases steadily with redshift. 

The filamentary structures and the connectivity of the more massive hence largest and latest formed haloes decrease over time \citep{Choi2010,Codis2018,Kraljic2020}. For our selected sample, the accretion tends to be more isotropic over time, with a tendency to become more spherical. Even though the filamentary accretion importance weakens with time, haloes with recent merger activity, still accrete at a higher rate than the rest of haloes (see the instantaneous logarithmic growth rate and the formation redshift presented in Fig.~\ref{fig:rate_accretion} in the Appendix section \ref{sec:appendix}).

Different physical processes intervene in the evolution in the accretion of mass and in the merger rate: the initial mass and statistics of the primordial density field \citep[][]{Bond1991}, mass and kinematic of subhaloes \citep[e.g.][]{Zhao2003}, and tidal forces \citep{Lapi2011}, which are possibly conditioned by dark energy \citep[][]{Pace2019}. The filament ($\ell=2$) higher strength at higher redshifts can be explain due to the effect of satellite haloes closer to the main one during the late stage of evolution before virialisation \citep{Schimd2021}. 

\begin{figure}
\centering
\includegraphics[width=\columnwidth]{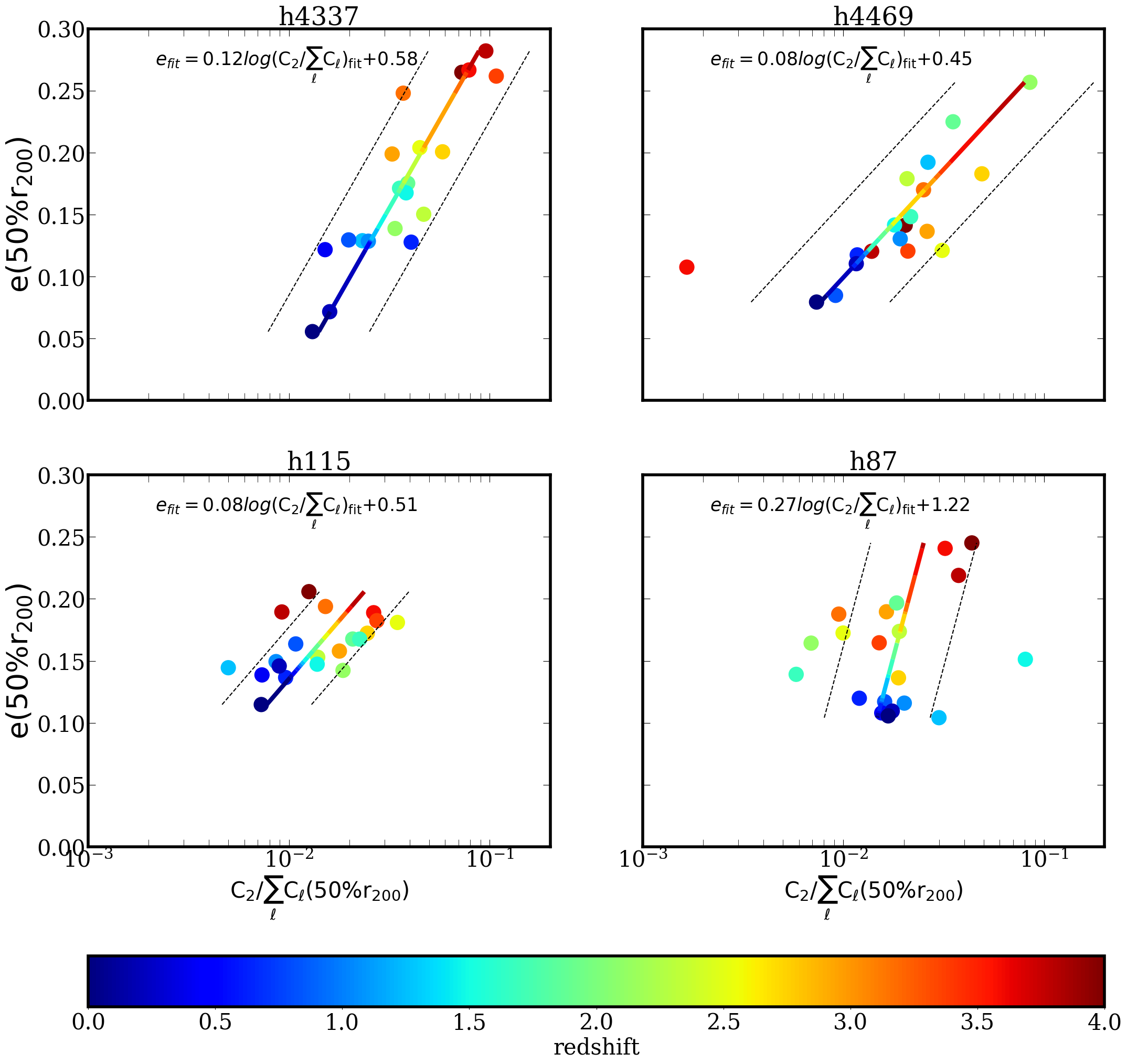}
\includegraphics[width=\columnwidth]{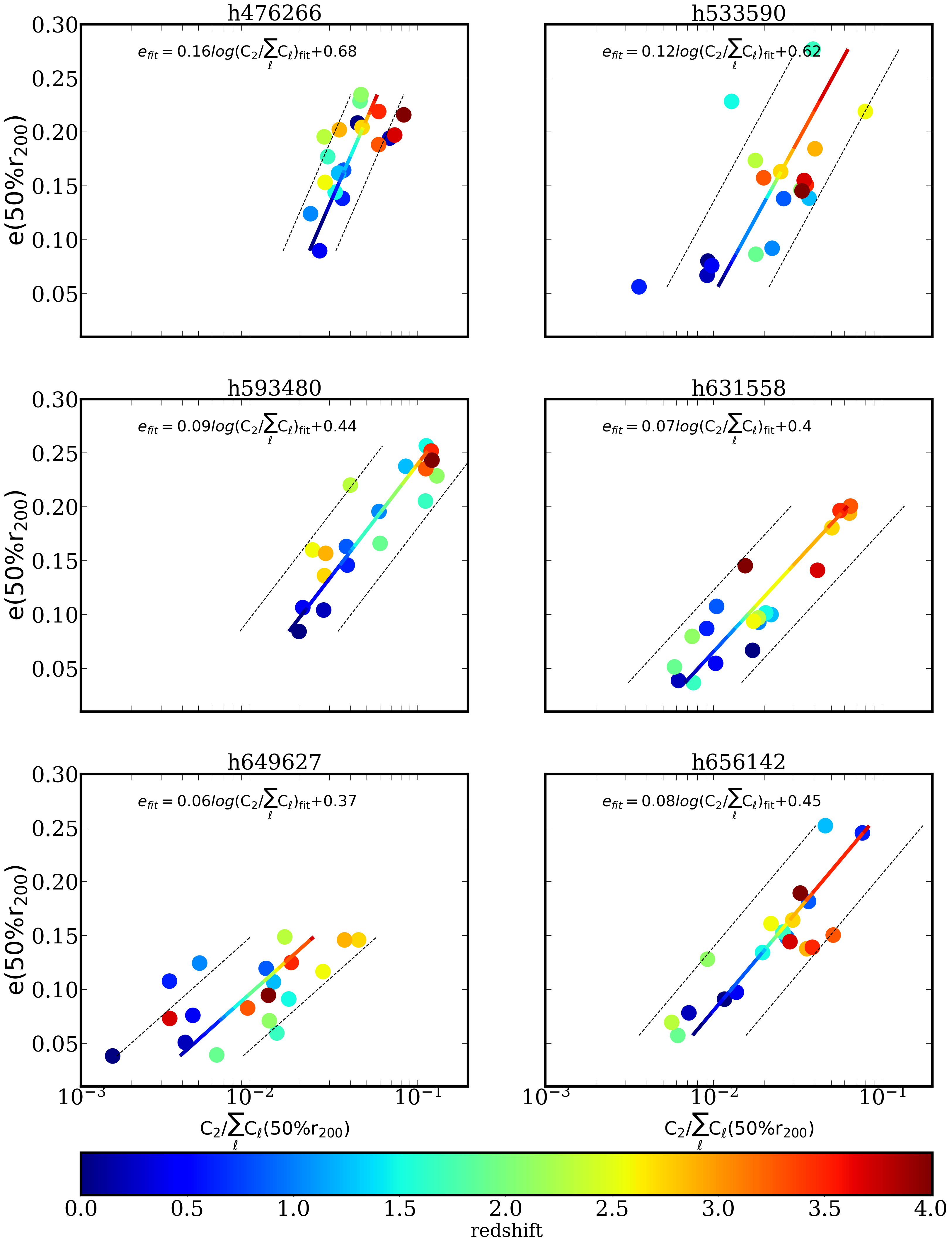}
\caption{The relation of quadrupole strength and ellipticity at $\mathrm{50 \% r_{200}}$. The accumulation of matter induce an increasing ellipticity in the direction of infalling matter. The filament strength decrease in recent time, with a subsequent less mergers events, which makes the haloes more spherical and less elliptic. A logarithmic regression fit is included coloured according to the median redshift in each bin of $e$, along its 1$\sigma$ dispersion (dashed black lines). Each panel shows the linear regression slope obtained.}
\label{fig:shape_vs_modes}
\end{figure}

In order to find a connection between filamentary accretion and ellipticity, we shift our analysis from $\mathrm{20 \% r_{200}}$ to outer regions $\mathrm{50 \% r_{200}}$, where the filamentary accretion still have a clear signal. In central regions, particle motions are dominated by other physical processes, with a subsequent loss in the quadrupole signal. Accordingly, in Fig.~\ref{fig:shape_vs_modes}, we inspect how the halo shape is influenced by the filament strength at $\mathrm{50 \% r_{200}}$. We perform a logarithmic regression between the ellipticity $e$ and $\mathrm{C_{2}/\sum _{\ell}C_{\ell}}$. The linear regression yields a positive slope for all haloes. The relation between ellipticity and filamentary accretion suggests that the accumulation of matter in a particular orientation increases the ellipticity of the halo. This connection also has a dependency on time, reflected by the color of the linear regression line in Fig.~\ref{fig:shape_vs_modes}. As the filament strength decreases steadily with redshift, the haloes became more spherical and less elliptical.

\section{Conclusions}
\label{sec:Conc}

In this work, we studied the properties of DM halo shapes and the interconnection with the halo assembly evolution. For this purpose, we study two samples from simulations with different subgrid physics implementations: a subsample of MW-like haloes from the highest resolution box of the {\sc  IllustrisTNG} Project (TNG50) and from the \cielo~zoom-in simulation. We investigate the evolution of the halo shape with redshift and the physical processes affecting them. Our main findings can be summarized as follow:
 
\begin{enumerate}

     \item{The concentration of the halo density profiles increases at lower redshifts and in more massive haloes. This increment in concentration, a product of both, the baryonic condensation in the inner regions and the relaxation of haloes, decreases when merger events happen. The evolution of halo size experience two different regimes before and after the turnaround point at $z \sim 2$. The MAH also shows the effects of mergers. Haloes with major mergers in recent times, reach afterward $\mathrm{z_{form,50}}$.}
     
     \item{We find that at more recent redshifts, haloes become less elliptical and in correspondence, more spherical in central regions. Also, haloes evolve to be more spherical/oblate in the inner regions and more triaxial in the outer ones. In the case of TNG50 haloes, this evolution is also present albeit weaker.}
    
    \item{For all analysed samples, we find that morphology tends to be more spherical as we go towards inner regions at $z=0$, with different paths for the shape parameter space between haloes. Focusing on $\mathrm{20 \% r_{200}}$ (where we expected the  halo morphology to maximize their changes \citep{Cataldi2020}), the evolution of the shapes shows a tendency toward more spherical configurations, although this evolution has a dependence on the merger force through time. In the case of TNG50 haloes, the path through the shape parameter space is more diverse and has a weaker tendency to sphericalization with time than \cielo{} haloes. In particular, h593480 shows a final oblate configuration at $z=0$. The effects of recent major accretion (h4469 and h476266) can be spot in the figures as a larger dispersion reaching lower redshifts.}

    \item{Exploring the halo assembly history can provide insight into the connection of mergers and halo shapes. We find that all haloes accrete matter with a dominant isotropic (i.e., monopole) accretion mode. The quadrupole mode (i.e. filaments) has the next dominant contribution in accretion, with a preferential direction. We find that the strength of the quadrupole mode decreases with radii and also with redshift, as haloes lose their connection to the cosmic web.}
    
    \item{We find a stronger connection between the strength of the filament and the degree of ellipticity of the halo shape at $\mathrm{50 \% r_{200}}$. The filaments with a given preferential direction, accumulate mass so that the halo axis becomes elongated in the same direction as the infalling matter. This results in haloes being more ellipticals. With the weakening of the preferential direction of accretion in recent times, the accretion of mass becomes more isotropic, with a subsequent transformation to more spherical shapes. We find that this connection can be well described as a logarithmic regression fit for all our halo selections.} 
    
\end{enumerate}

Our results show that the assembly history has a key role in understanding the resulting halo morphology at $z=0$. There is an interconnection between the halo shape driven by the cosmic web at outskirt radii and the assembly of baryons at inner regions. The shape evolution is an important ingredient to understand the halo assembly history as well as the merger history.

\section*{Acknowledgements}

PC and SP acknowledges partial support from MinCyT through BID PICT 2020 00582. PBT acknowledges partial support from Fondecyt Regular 20201200703 and ANID BASAL project ACE210002 (Chile). This project has received funding from the European Union’s Horizon 2020 Research and Innovation Programme under the Marie Skłodowska-Curie grant agreement No 734374 (LACEGAL) and the GALNET Network (ANID, Chile). The \cielo~ Project was run in Marenostrum (Barcelona Supercomputer Centre, Spain), Ladgerda (IA, PUC) and the National Laboratory for High Performance Computing (NLHPC, Chile). We thank the Ministerio de Ciencia e Innovación (Spain) for financial support under Project grants PGC2018-094975-C21 and PID2021-122603NB-C21. MCA acknowledges financial support from the Seal of Excellence @UNIPD 2020 program under the ACROGAL project.

\section*{Data Availability}
The data underlying this article will be shared on reasonable request to the corresponding authors.
\bibliographystyle{mnras}
\bibliography{bibliography} 
\newpage
\appendix

\section{Extended analysis of halo shape evolution}
\label{sec:appendix}

\setcounter{table}{0}
\renewcommand{\thetable}{A\arabic{table}}
\setcounter{figure}{0}
\renewcommand{\thefigure}{A\arabic{figure}} 

\begin{figure}
\centering
\includegraphics[width=0.51\columnwidth]{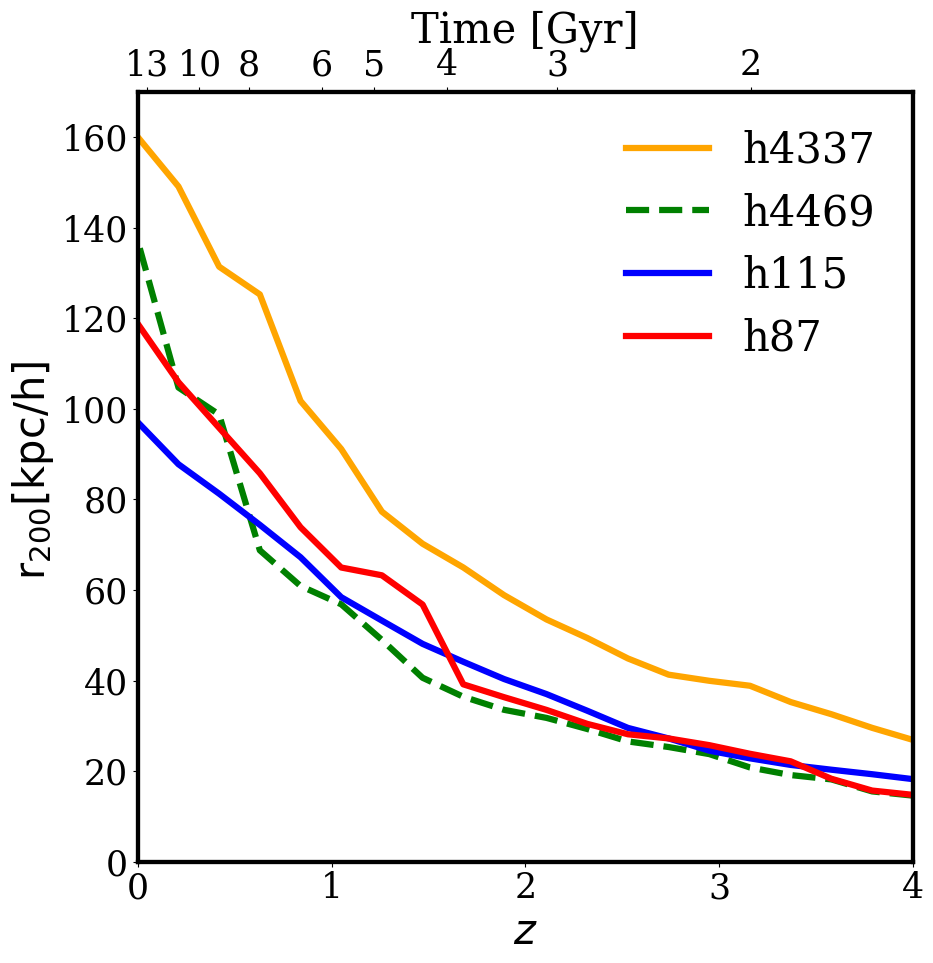}
\includegraphics[width=0.455\columnwidth]{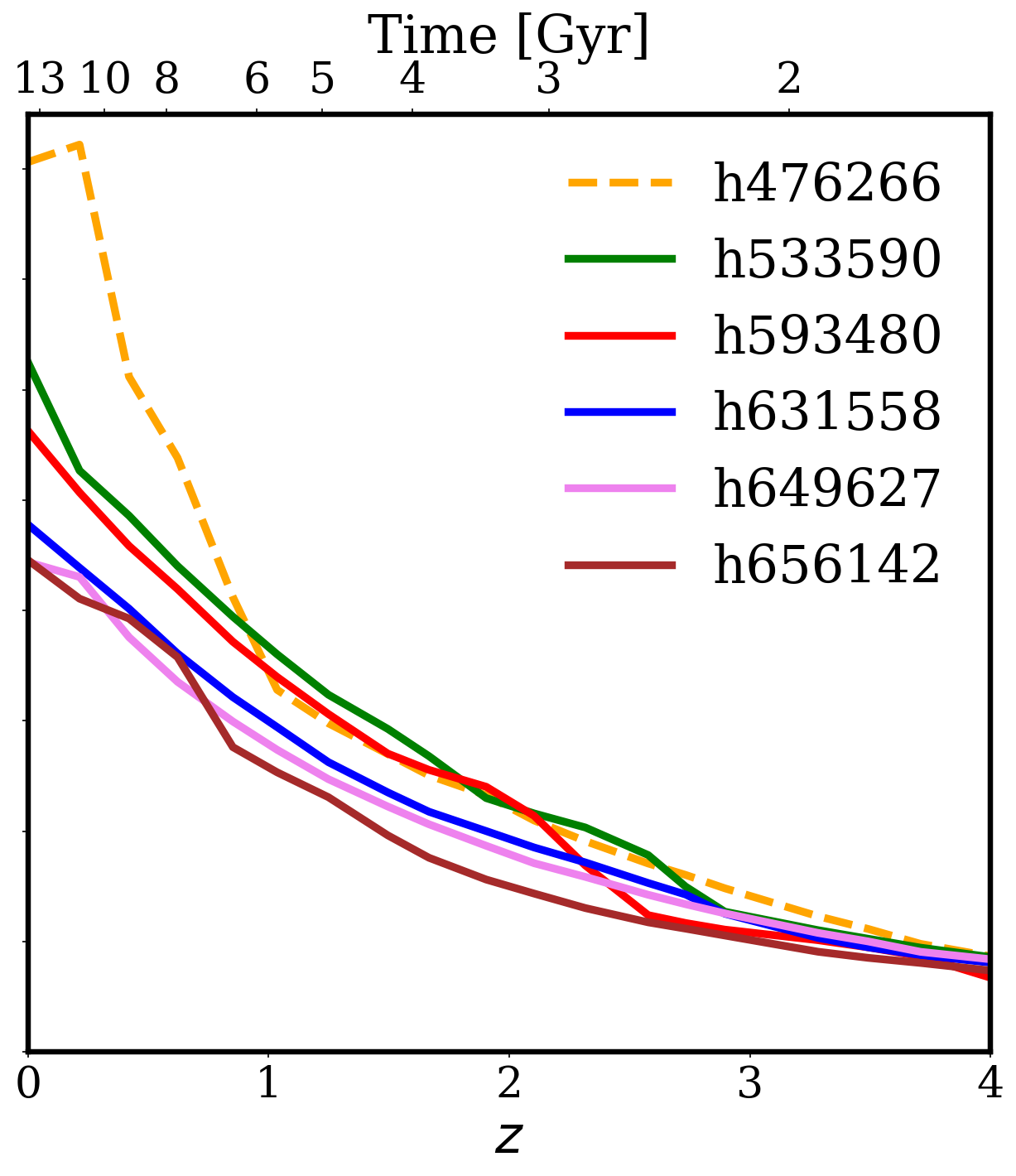}
\caption{The cosmic evolution of the virial radius, $\mathrm{r_{200}}$, for the DM haloes of \cielo{} (\textit{left panel}) and TNG50 (\textit{right panel}). Dashed lines indicate the haloes with recent mergers activity. }
\label{fig:r_vs_z}
\end{figure}

In Fig.~\ref{fig:r_vs_z} we show the evolution of the virial radius $\mathrm{r_{200}}$. We observe two different regimes in the evolution. First, a slow increase of halo size over time, up to $z=2$, follow by an acceleration in the increasing of $\mathrm{r_{200}}$. The effects of recent mergers activity (h476266) or major accretion (h4469) can be spot with a sudden increment at late times of the size of the halo. 

\begin{figure}
\centering
\includegraphics[width=0.52\columnwidth]{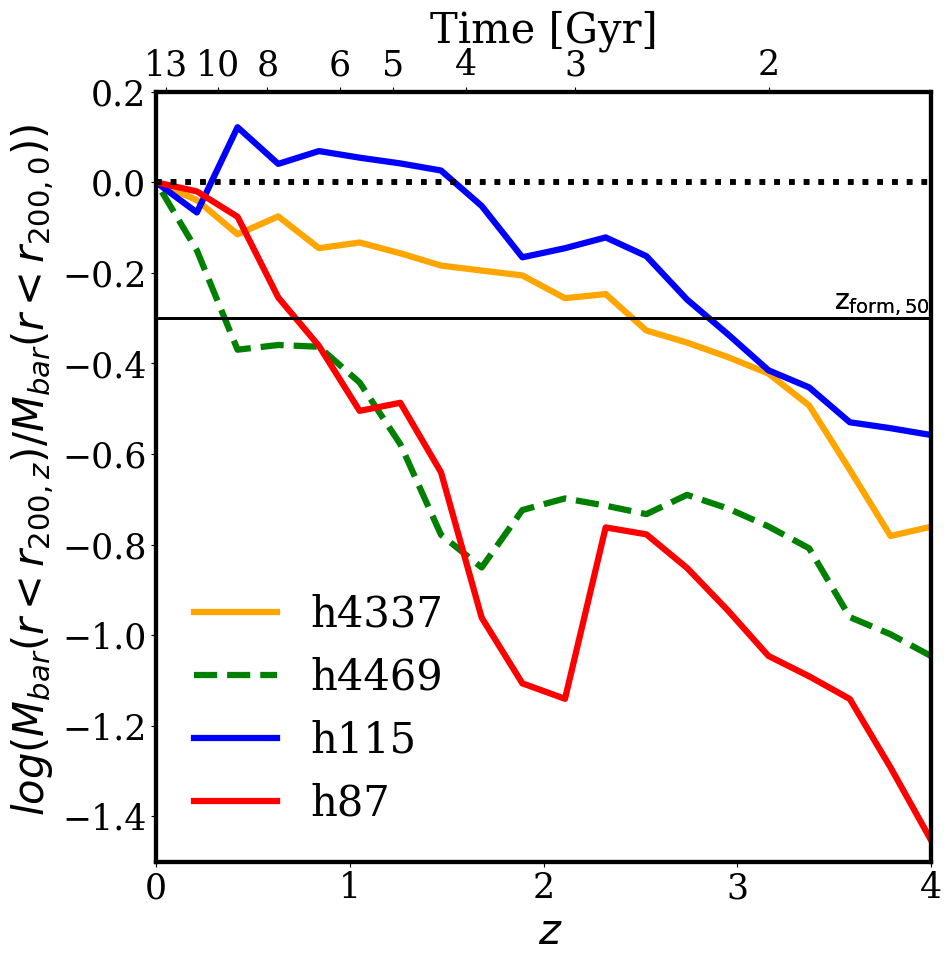}
\includegraphics[width=0.46\columnwidth]{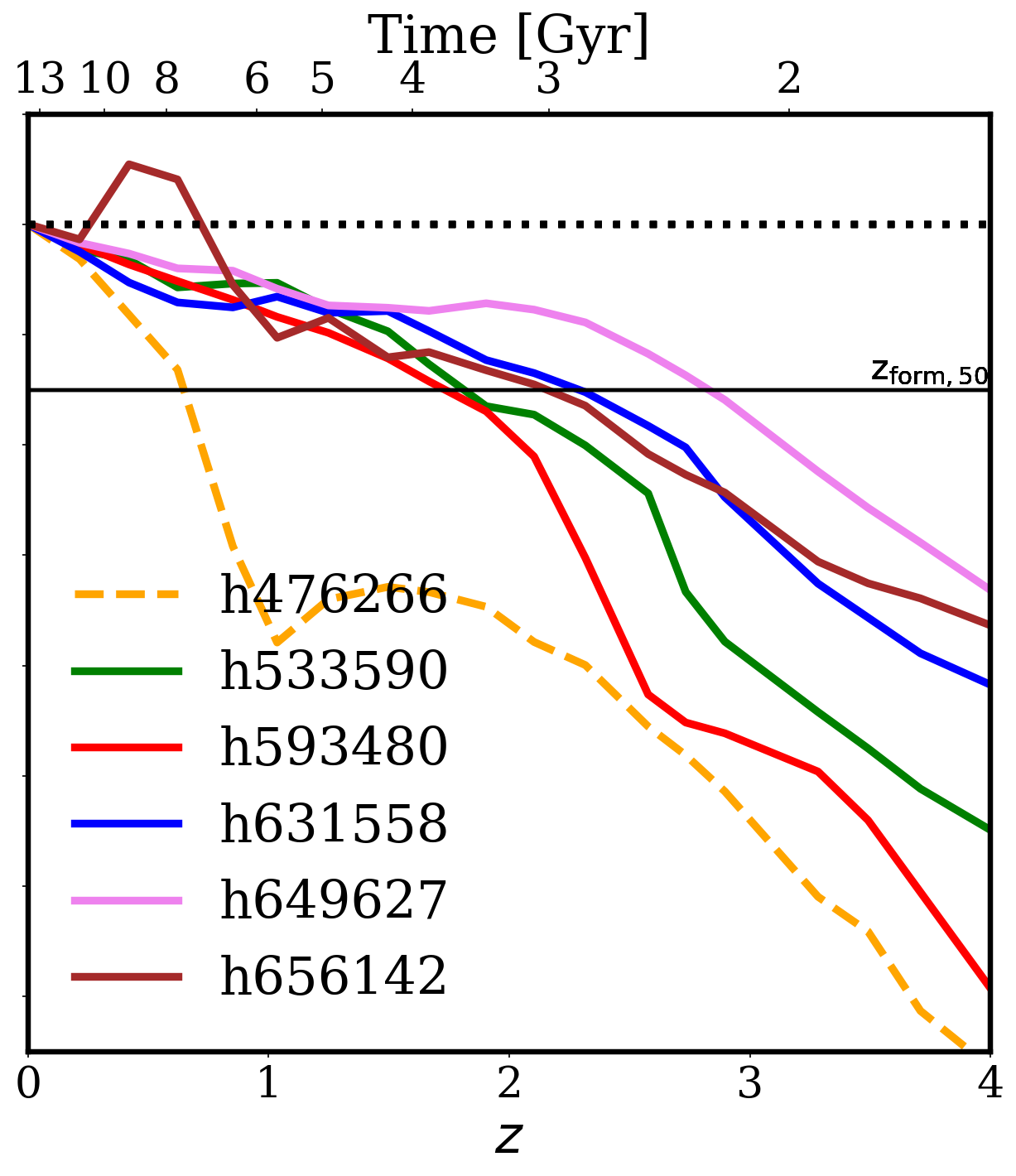}
\caption{The baryonic MAH of the haloes vs redshift $z$, in \cielo{} \textit{(left panel)} and TNG50 \textit{(right panel)} haloes. Dashed lines indicate the haloes with no constraints in mergers activity. Black horizontal line ($\mathrm{z_{form,50}}$) in the baryonic MAH panels is used as reference to estimate the formation time of the haloes, as the redshift at which the mass of the halo reach half of their mass at $z =0$.}
\label{fig:MAH_b}
\end{figure}

We also study he evolution of baryonic MAH for the selected haloes in Fig.~\ref{fig:MAH_b}. The accretion of baryons (stars and gas) follow the same trend as DM MAH. Inspecting Fig.~\ref{fig:MAH_b}, the baryons have a MAH slope which is less monotonically increasing than DM MAH in Fig.~\ref{fig:MAH}, product of the star formation and complex gas dynamics within each galaxy.  

By computing the mass growth of halos over cosmic time, we can calculate different proxies of their mass assembly history. Several studies describe the instantaneous mass accretion rate \citep[e.g.][]{RodriguezPuebla2016,Gouin2021,Montero-Dorta2021} as,

\begin{equation}
\mathrm{\frac{dM_{200}}{dt}=\frac{M_{200}(t+dt)-M_{200}(t)}{dt}}
\end{equation}

\begin{figure}
\centering
\includegraphics[width=0.51\columnwidth]{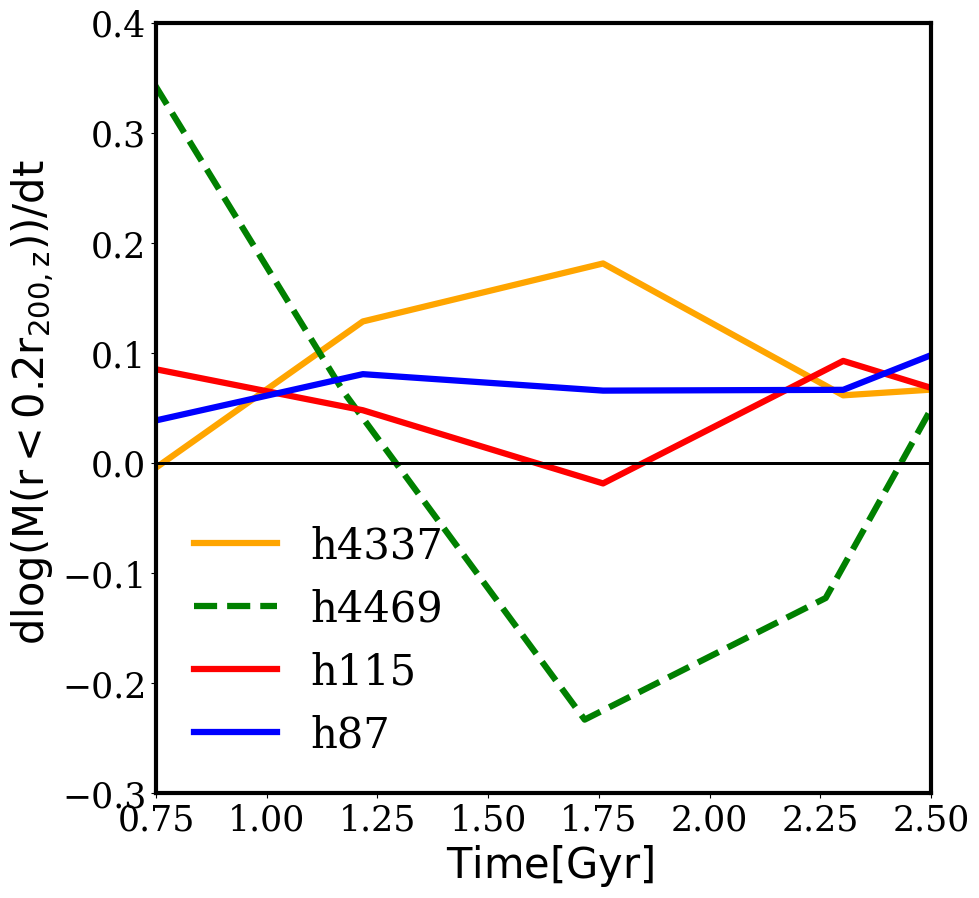}
\includegraphics[width=0.465\columnwidth]{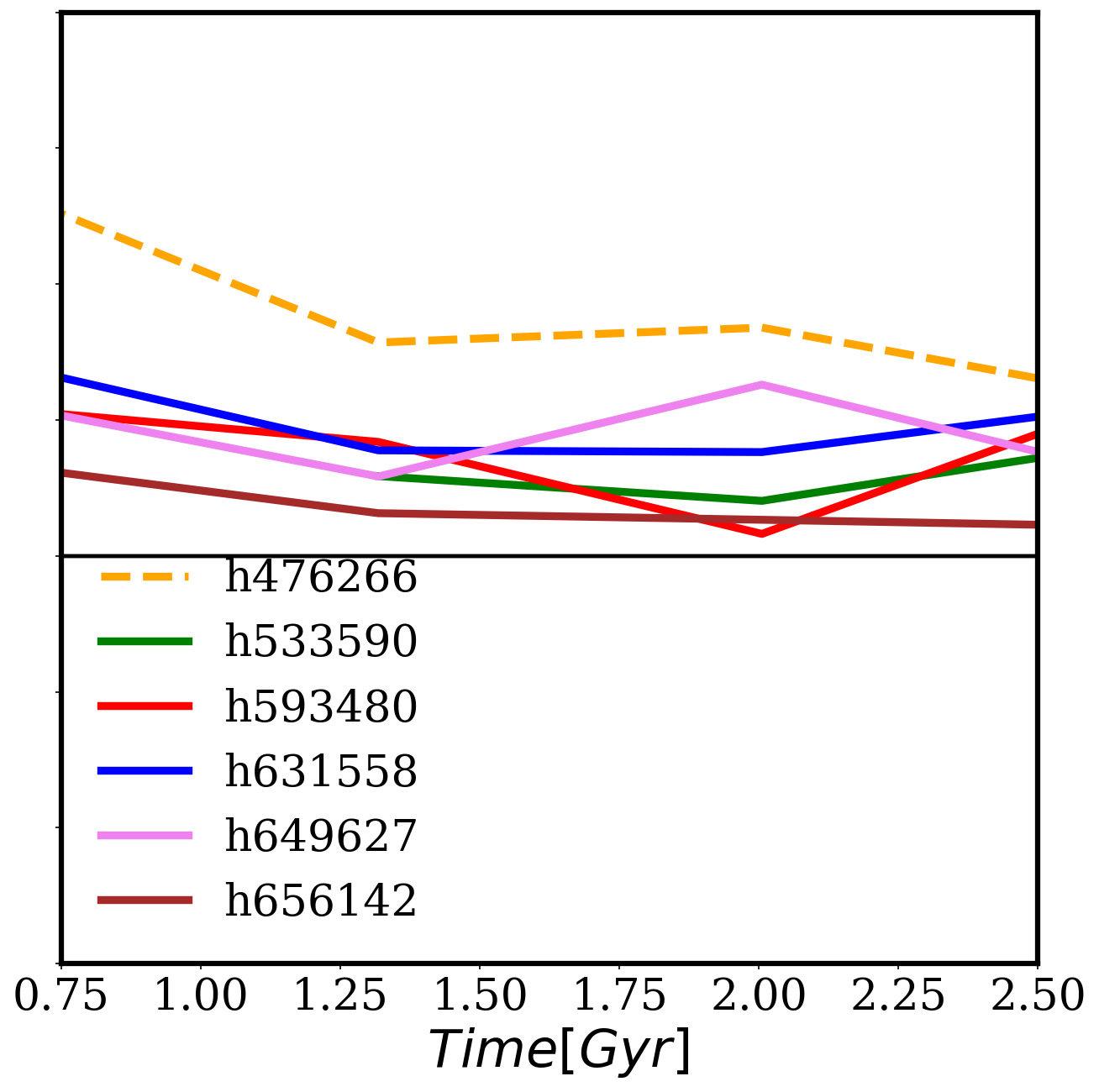}
\caption{The instantaneous halo growth rate for \cielo{} \textit{(left panel)} and TNG50 haloes \textit{(right panel)} as a function of lookback time.}
\label{fig:rate_accretion}
\end{figure}

In Fig.~\ref{fig:rate_accretion} we estimated the logarithmic halo growth rate. We choose for this, to compute the mass at $\mathrm{20\% \, r_{200}}$, where we studied the changes in morphologies.

 In Tab.~\ref{Tab:time_formation} we present the instantaneous logarithmic growth rate $\mathrm{dlog(M)/dt_{z \sim 0}}$ at $z=0$, together with the formation redshift at 70$\%$ ,$\mathrm{z_{form,70}}$, defined as the redshift which the mass of the halo main progenitor at $\mathrm{20\%  \, r_{200}}$ is equal to 70$\%$ of the mass enclosed at the same radius at $z=0$. Haloes with recent mergers accretion (h4469) and  merger activity (h476266) are the last to reach $70 \%$ of the mass at $z=0$ and consequently these are the haloes with greater instantaneous growth rate at $z=0$.

\begin{table}
	\centering
	\caption{The formation redshift at 70$\%$ ,$\mathrm{z_{form,70}}$, and the instantaneous logarithmic growth rate $\mathrm{dlog(M)/dt_{z \sim 0}}$ at $z=0$ for the selected haloes.}
	\label{Tab:time_formation}
	\begin{tabular}{lcc} 
		\hline
		\cielo{} & $\mathrm{z_{form,70}}$ & $\mathrm{dlog(M)/dt_{z \sim 0}}$ \\
		\hline
		h4337 & 0.75 & -0.02 \\
		\textbf{h4469} & 0.44 & 0.40  \\
		h87 & 0.88 & 0.09  \\
		h115  & 0.55 & 0.03 \\
		\hline
    	TNG50 & $\mathrm{z_{form,70}}$ & $\mathrm{dlog(M)/dt_{z \sim 0}}$  \\
        \hline
		\textbf{h476266} & 0.48 & 0.27 \\
		h533590  & 0.82 & 0.05  \\
		h593480  & 0.68 & 0.11   \\
		h631558  & 0.70 & 0.14  \\
		h649627  & 0.85 & 0.11 \\
		h656142  & 0.62 & 0.06  \\  
        \hline
	\end{tabular}
\end{table}

\bsp	
\label{lastpage}
\end{document}